\documentclass[a4paper,11pt]{paper}

\usepackage[english]{babel}
\usepackage[utf8]{inputenc}
\usepackage[T1]{fontenc}

\usepackage[a4paper,top=2cm,bottom=2cm,left=3cm,right=3cm,marginparwidth=1.75cm]{geometry}

\usepackage{nomencl}
\usepackage{etoolbox}
\usepackage{multirow}
\usepackage{multicol}
\usepackage{threeparttable}
\usepackage{siunitx}
\usepackage{amssymb}
\usepackage{amsmath}
\usepackage{tikz}
\usetikzlibrary{fit}
\usepackage{float}
\usepackage{algorithm}
\usepackage{algpseudocode}
\usepackage{empheq}
\usepackage{booktabs}
\usepackage{graphicx}
\usepackage{subfigure}
\usepackage{xcolor}
\usepackage{colortbl}
\usepackage{adjustbox}
\usepackage[colorlinks=true, allcolors=blue]{hyperref}
\usepackage[authoryear,round]{natbib}
\usepackage{notoccite}

\definecolor{tcd_blue}{RGB}{5, 105, 185}

\title{Sequential Monte Carlo Squared for Online Inference in Stochastic Epidemic Models}

\author{Dhorasso Temfack and  Jason Wyse\\
\textit{School of Computer Science and Statistics, Trinity College Dublin, Ireland}}

\begin{document}

\maketitle

\begin{abstract}
Effective epidemic modeling and surveillance require computationally efficient methods that can continuously update parameter estimates as new data becomes available. This paper explores the application of an online variant of Sequential Monte Carlo Squared (O-SMC$^2$) to the stochastic Susceptible-Exposed-Infectious-Removed (SEIR) model for real-time epidemic tracking. The advantage of O-SMC$^2$ lies in its ability to update parameter estimates using a particle Metropolis-Hastings kernel by only utilizing a fixed window of recent observations. This feature enables timely parameter updates and significantly enhances computational efficiency compared to standard SMC$^2$, which by comparison, requires processing of all past observations.  First, we demonstrate the efficiency of O-SMC$^2$ on simulated epidemic data, where both the true parameter values and the observation process are known. We then make an application to a real-world COVID-19 dataset from Ireland, successfully tracking the epidemic and estimating a time-dependent reproduction number of the disease. Our results show that O-SMC$^2$ provides highly accurate online estimates of both static and dynamic epidemiological parameters while substantially reducing computational cost. These findings highlight the potential of O-SMC$^2$ for real-time epidemic monitoring and supporting adaptive public health interventions.
\end{abstract}



\vspace{1em}

\section{Introduction}
Disease modeling plays a crucial role in public health by providing insights into the mechanisms behind the spread of infectious diseases. Such understanding is vital for designing effective intervention strategies and supporting evidence-based policy-making. As the global landscape of infectious diseases evolves and the risk of novel a pandemic events like COVID-19 increases \citep{marani2021intensity}, there is a pressing need for robust modeling frameworks that can adapt to changes in the underlying disease dynamics, progressively update parameter values as new information becomes available, and deliver accurate assessments and predictions of epidemic trajectories \citep{Birrell2017}.

Traditionally, parameter inference in epidemic models has been approached using both frequentist and Bayesian methods. Frequentist techniques such as maximum likelihood estimation \citep{althaus2014estimating} and nonlinear least squares \citep{aloke2023parameters} have been widely applied to deterministic models based on ordinary differential equations due to their simplicity and computational efficiency. Bayesian approaches, have gained popularity for their ability to incorporate prior knowledge and quantify uncertainty. In this context, Markov chain Monte Carlo (MCMC) methods remain a popular choice for sampling from posterior distributions of model parameters \citep{o1999bayesian,jewell2009bayesian, wang2022bayes}. Recent advances in probabilistic programming tools like Stan \citep{carpenter2017stan, grinsztajn2021bayesian} and specialized toolboxes such as \texttt{quantdiffforecast} \citep{chowell2024parameter} and \texttt{Bayesianfitforecast} \citep{karami2024bayesianfitforecast} have made Bayesian workflows more accessible and flexible in epidemiological modeling. However, the computational demands of MCMC limit its utility for real-time applications. This is because incorporating incrementally observed new data points typically necessitates re-running the entire inference process \citep{Birrell2017}. When the likelihood function is intractable but simulation from the generative model is feasible, likelihood-free approaches such as Approximate Bayesian Computation (ABC) are commonly used \citep{kypraios2017tutorial}.  While ABC methods offer flexibility and an intuitive framework, their performance relies heavily on a user chosen set of summary statistics to measure the distance between simulated and observed data. Poorly chosen, these can detrimentally impact performance. Standard ABC lacks native mechanisms for sequential parameter updating as new data arrive incrementally \citep{minter2019approximate}. Sequential adaptations of ABC are possible \citep{toni2009approximate}, however these still depend heavily on user chosen summary statistics, creating challenges in complex modelling scenarios \citep{han2025sequential}. 

We adopt an approach which gets around many of the limitations outlined above, and enables sequential and probabilistically well calibrated inference. Sequential Monte Carlo (SMC) methods, also known as particle filtering, provide a natural framework for real-time state and parameter estimation by sequentially updating a posterior distribution as new data becomes available \citep{Doucet2001}. These methods have seen growing application in epidemiology, with early studies demonstrating their effectiveness in calibrating nonlinear compartmental models of infectious disease dynamics \citep{yang2014comparison, camacho2015temporal}. For example, \cite{welding2019real} applied a sequential Bayesian framework to the 2001 UK Foot-and-Mouth epidemic, using MCMC-based rejuvenation steps to mitigate particle degeneracy and enable real-time outbreak analysis. Similarly, \cite{birrell2020efficient} developed an age-structured SEIR model on simulated influenza outbreak data, demonstrating efficient real-time inference by integrating multiple data sources and reducing computational costs compared to standard MCMC.

While SMC methods are well-suited for sequential state estimation, they are not directly able to facilitate the estimation of static parameters \citep{kantas2015particle}. Hybrid techniques such as particle MCMC (PMCMC) \citep{Andrieu2010} address this limitation by combining SMC with MCMC. PMCMC generates unbiased likelihood estimates through particle filtering, enabling exact posterior sampling via traditional MCMC methods. However, these methods remain computationally demanding since each MCMC iteration requires a full SMC run, limiting their suitability for real-time applications \citep{Dureau2013, funk2018realtime, endo2019introduction, cazelles2021dynamics}. In their recent study, \cite{storvik2023sequential} propose an SMC framework for inferring a time-varying reproduction number of COVID-19 in Norway.  Their approach sequentially estimates hyperparameters controlling the reproduction number by relying on a set of sufficient statistics. However, this approach assumes that the sufficient statistics-based filters are analytically tractable, a condition that only holds for particular types of dynamical systems.

A promising alternative is the SMC$^2$ algorithm \citep{Chopin2013, Jacob2015, rosato2023log}, which integrates nested particle filters to simultaneously estimate model states and parameters, improving both accuracy and adaptability. A key step in SMC$^2$ is the so-called ``rejuvenation'' using PMCMC kernels to prevent particle impoverishment. Nevertheless, this step conventionally requires computing the likelihood over all past observations, posing challenges for truly online or streaming data applications.

To address this limitation, \cite{vieira2018bayesian} introduced a computationally efficient modification that evaluates the PMCMC over a fixed-size moving window of recent observations. In this study, we apply and evaluate this fixed-window SMC$^2$ approach, referred to as online SMC$^2$ (O-SMC$^2$), within the context of epidemic modeling.  While O-SMC$^2$ is an established algorithm, its strengths for disease modeling have not been explored. Our contribution is its tailored implementation and comprehensive assessment for sequential state and parameter inference in stochastic SEIR models under real-time constraints. This is an area of growing practical importance for epidemic response planning. Compared to standard SMC$^2$, the O-SMC$^2$ framework substantially reduces computational demands while preserving the capacity for accurate sequential estimation. This efficiency gain is particularly relevant for fast-moving outbreaks, where timely estimates of key epidemiological parameters such as the time-varying effective reproduction number and the infectious period are critical for guiding public health interventions.

The remainder of this paper is structured as follows. Section \ref{sec2} describes the state-space framework for the SEIR compartmental model, provides a brief overview of the SMC$^2$/O-SMC$^2$ methodology, and discusses parameter identifiability. In Section \ref{sec3}, we apply the O-SMC$^2$ algorithm to both simulated data and real-world COVID-19 data from Ireland, demonstrating its potential for real-time epidemiological surveillance and inference of the time-dependent reproduction number. Finally, Section \ref{sec4} concludes the paper by summarizing the main findings and discussing potential challenges and limitations of our approach.

\section{Compartmental model and sequential inference}\label{sec2}

\subsection{Stochastic SEIR model}
\begin{figure}[H]
        \centering
   \includegraphics[scale=1]{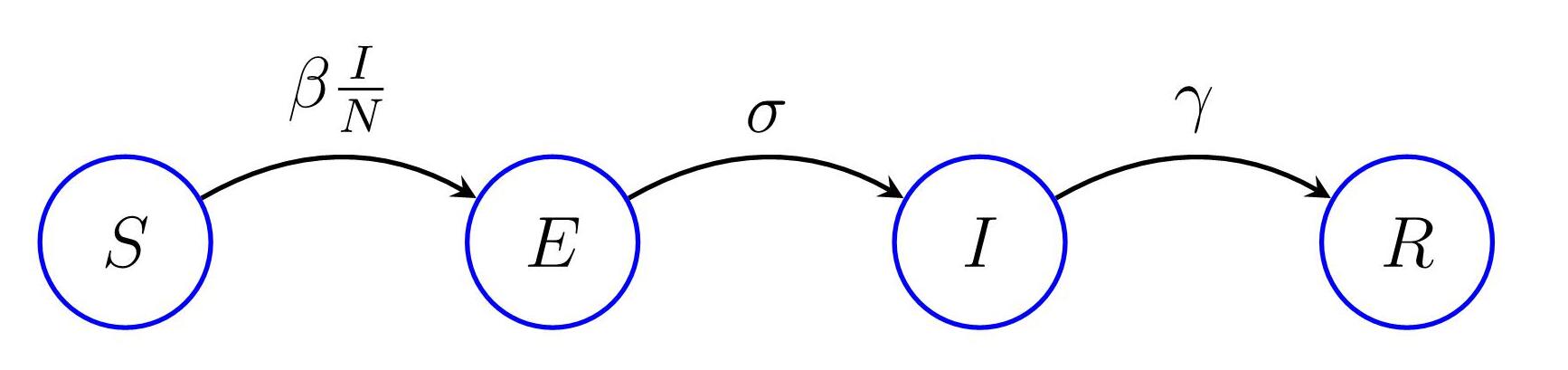}
    \caption{\footnotesize{ Graphical representation of a SEIR model}. Each arrow expresses the rate of transition between the compartments. $\beta$ is the disease transmission rate, $1/\sigma$ is the mean incubation period, and $1/\gamma$ is the mean infectious period.}
    \label{figure:Fig1}
\end{figure}

Compartmental models for infectious diseases, which had their conceptual origins in the early to mid $20^{th}$ century, attempt to describe a disease's progression and spread through individuals moving through disease stages over time. The stochastic SEIR model serves as the foundation for our investigation. Consider the time interval $(t,t+\delta t]$, where $\delta t$ represents the time between observations.  If we assume that the time spent by an individual in a compartment is exponentially distributed with a rate $\lambda $, then the number of individuals leaving a compartment during the time step $\delta t$ can be seen as a Binomial random variable with success probability $p=1-e^{-\lambda\delta t}$ (this parameterization ensures a success probability between $0$ and $1$, see \cite{lekone2006statistical}). The discrete-time  SEIR stochastic  model is specified by:
\begin{align} \label{stoseir}
S_{t+ \delta t} &= S_{t} - Y_{SE}(t), & Y_{SE}(t)& \sim\mathrm{Bin}\left(S_{t}, 1-e^{-\beta \frac{ I_{t}}{N} \delta t}\right)  \notag\\
E_{t+ \delta t} &= E_{t} + Y_{SE}(t)-Y_{EI}(t), &Y_{EI}(t)& \sim\mathrm{Bin}\left(E_{t}, 1-e^{-\sigma \delta t}\right)\\
I_{t+ \delta t} &= I_{t} + Y_{EI}(t)-  Y_{IR}(t), &Y_{IR}(t)& \sim\mathrm{Bin}\left(I_{t}, 1-e^{-\gamma  \delta t}\right) \notag\\
R_{t+ \delta t} &= R_{t} + Y_{IR}(t). \notag
\end{align}
Here, $S_t$, $E_t$, $I_t$, and $R_t$ represent the compartments of susceptible, exposed (infected but not yet infectious), infectious, and removed individuals, respectively, at time $t$. Susceptible individuals, upon contact with infectious individuals, transition to the exposed class at a rate $\beta$ (which can be time-dependent or constant). Exposed individuals become infectious at a rate $\sigma$, and infectious individuals recover from the disease at a rate $\gamma$ (Figure~\ref{figure:Fig1}). The population is assumed to have a constant size $N = S_t + E_t + I_t + R_t$ at each time point.

\subsection{State-space framework}

\begin{figure}[H]
            \centering     
  \includegraphics[scale=1]{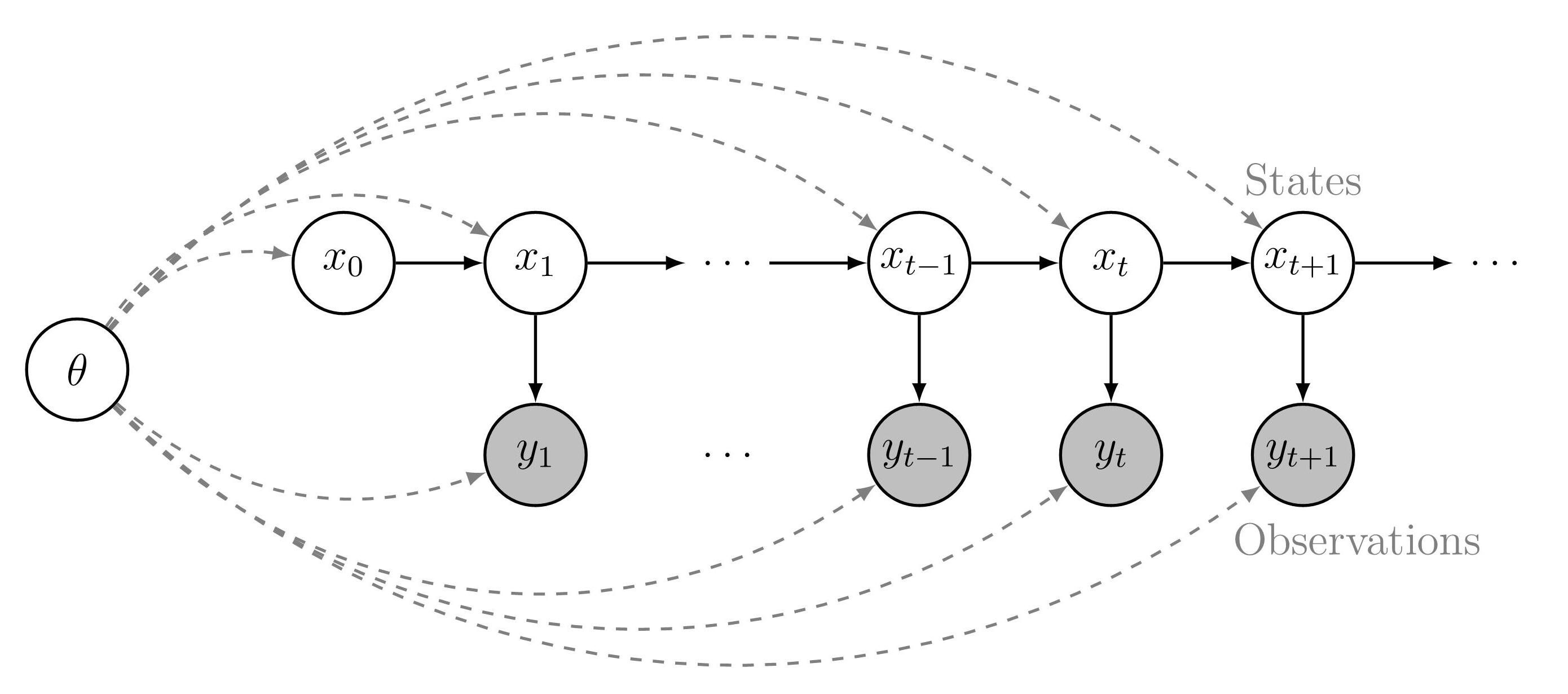}
\caption{\footnotesize { Graphical representation of a state-space model.} Black arrows show variable dependencies, and dashed gray lines represent the dependence on $\theta$.}
    \label{figure:Fig2}
\end{figure}

In the context of the SEIR model, the system can be formulated as a state-space model (SSM), which consists of two stochastic processes: one governing the evolution of the disease within the population, and the other describing the observation process (Figure~\ref{figure:Fig2}). The state-space model is defined for time steps $t = 1, \ldots, T$ as follows:
\begin{align}
    & x_0 \sim f(x_0 |\theta),  &  \rhd & \qquad \text{Initial state} \label{eq:initial_state} \\
    & x_t | x_{t-1} \sim f(x_t | x_{t-1}, \theta),  & \rhd & \qquad \text{State transition process} \label{eq:state_process} \\
    & y_t | x_t \sim g(y_t | x_t, \theta).  & \rhd& \qquad \text{Observation process} \label{eq:observation_process}
\end{align}
Here, $x_t$ denotes the vector of latent state variables (state variable in the classical SEIR model), which evolve according to a Markov process with transition distribution $f(x_t | x_{t-1}, \theta)$ given by the model \eqref{stoseir}. The observation $y_t$ is assumed to be probabilistically related to the hidden state $x_t$ via the distribution $g(y_t | x_t, \theta)$. The parameter vector $\theta$ encompasses all static epidemiological parameters of the model such as  $\beta$, $\sigma$ and  $\gamma$, governing both the state transitions and the observation process. Throughout the study,  $\theta$ refers collectively to all state space model and observation process parameters.

State-space models are particularly useful for modeling complex dynamical systems such as disease transmission where the true states are not directly observable, and observations are subject to noise or other forms of uncertainty. By explicitly modeling the hidden states and their relationship to observed data, we can perform inference on both the state variables and the model parameters.

\subsection{State and Parameter inference}
In this section, we briefly describe the procedure for sequential parameter estimation using the SMC$^2$ scheme. Let $x_{0:t}$ \footnote{For any sequence $\{y_t\}_{t\geq 0}$, we denote $z_{i:j} = \{z_i, z_{i+1}, \ldots, z_j\}$ and $z^{1: N}_j$ (resp. $z^{1: N}_{i:j}$) is a set of $N$ random variables $z^l_j$ (resp. $z^{l}_{i:j}$), for $ l=1,..., N$.} represent the sequence of latent states (in our case the SEIR compartment counts) up to time $t$, and let $y_{1:t}$ represent the observations (e.g., daily reported cases). The goal is to construct a sequence of posterior distributions $ \{p(\theta | y_{1:t}),t = 1, \ldots, T \} $ that reflect the evolution of parameter estimates as new observations become available. Bayes' rule provides the basis for updating the posterior distribution as follows:
\begin{align}
 p(\theta| y_{1:t})   \propto p(\theta| y_{1:t-1})p(y_t | y_{1:t-1}, \theta),
\end{align}
which suggests a sequential importance sampling approach. This approach involves reweighting a set of parameter particles based on the incremental likelihood $p(y_t | y_{1:t-1}, \theta)$.  The iterated batch importance sampling (IBIS) method \citep{chopin2002sequential} employs this technique with an MCMC step to rejuvenate parameter samples. However, for state-space models, the incremental likelihood is generally intractable. To address this issue, we can replace $p(y_t | y_{1:t-1}, \theta)$ with an unbiased estimate derived from running a particle filter, targeting the filtering distribution $p(x_t | y_{1:t}, \theta) $ for each of the $\theta$-particles. This approach is known as  SMC$^2$ due to the use of nested filters. For a comprehensive discussion on this approach, including formal justifications, refer to \citep{Chopin2013}.

\subsubsection{Particle filtering}

For a given value of the model parameters $\theta$, the latent variables $x_{t}$ at each time point, can be inferred from the observations $y_{1:t}$ using the filtering distribution $ p(x_{t} | y_{1:t}, \theta)$. Particle filtering is a technique that approximates this filtering distribution by sequentially generating a set of particles with associated weights $\{x_t^{i}, w_t^{i}\}_{i=1}^{N_x}$, via importance sampling and resampling. Hence, for a known parameter value $\theta$, the posterior distribution of the latent states is represented as a Dirac delta mixture:
\begin{align}
   p(x_t |y_{1: t},\theta)\approx\dfrac{1}{\sum_{j=1}^{N_x} w_t^{j}}\sum_{i=1}^{N_x} w_t^{i} \delta_{x_t^{i}}(x_t) . 
\end{align}
Here, $N_x$ denotes the number of  $x$-particles, and  $w^{i}_{t}:=w_{t}^{i}(x_{t-1}^{i},x_t^{i})$ represents the importance weight for each particle $x_t^{i}$. 
The weight update rule at time $t$ takes into account both the likelihood of the new observation $y_{t}$ given the current particle state $x_{t}^{i}$, and the transition probability between successive states:
\begin{align} \label{wt2}
    w_{t}^{i} = w_{t-1}^{i}\dfrac{g(y_{t} | x_{t}^{i},\theta) f(x_{t}^{i} | x_{t-1}^{i},\theta)}{q(x_{t}^{i} | x_{t-1}^{i}, y_{t}, \theta)}, \quad  i = 1, \ldots, N_x,
\end{align}
One notable particle filtering method is the Bootstrap particle filter (BPF) \citep{gordon1993novel}, which simplifies weight calculations by setting the importance distribution equal to the state transition distribution, i.e,
$q(x_{t} | x_{t-1}, y_{t}, \theta)=f(x_{t} | x_{t-1},\theta)$. Many filtering algorithms have been developed to improve efficiency \citep{pitt1999filtering}, but for this study, we focus on the BPF. The general procedure for the Bootstrap Particle Filter is shown in Algorithm \ref{pf}.

\begin{algorithm}
\caption{Bootstrap Particle Filter (BPF) for a given parameter $\theta$}\label{pf}
Operations involving index $i$ must be performed for $i = 1,..., N_x$.

The indices $a_{t}^{1:N_x}$ define the ancestral particles at time $t$ after the resampling.

\textbf{Inputs:} Observation: $y_{1:T}$, Number of particles: $N_x$, Inital state distribution: $f(x_{0}|\theta)$, Parameter vector: $\theta$.

\textbf{Output:}  Particle set :$\left\{x^{i}_{0:t}, w^{i}_{0:t}\right\}_{i=1}^{N_x}$

\hrulefill 
\begin{algorithmic}[1]
    \State Sample initial  particles : $x^{i}_0 \sim f( x_0|\theta)$
    \State Compute weights:  $w^{i}_0 = 1$, \quad $W^{i}_0 = w^{i}_0/\sum_{j=1}^{N_x}w^{j}_0$ 
    \For{$t=1$ to $T$}
        \State\label{res1} Sample new indices: $a_{t}^{1:N_x}\sim\text{Resample}(W_{t-1}^{1: N_x})$
        \State Propagate states $x_t^{i} \sim f( |x_{t-1}^{a_{t}^{i}},\theta)$
        \State Compute new weights $w_t^{i}= g(y_t |x_t^{i},\theta)$
        , \quad $W^{i}_{t} = w^{i}_{t}/\sum_{j=1}^{N_x}w^{j}_{t}$ 
    \EndFor
\end{algorithmic}
\end{algorithm}

The resampling step in Algorithm \ref{pf}, referred to in step \ref{res1}, aims to mitigate particle degeneracy\footnote{Particle degeneracy refers to the situation where, after many iterations, only a small number of particles dominate and retain a large proportion of the weight.} by discarding particles with lower weights and duplicating those with higher weights. After resampling, we reset all weights to $w^{i}_{t-1}=1$. Various resampling schemes exist as described in \citet{doucet2009tutorial}. Unless stated otherwise, we use stratified resampling in all our implementations due to its efficiency and lower computational cost \citep{SHEINSON201421, vieira2018bayesian}.  In Algorithm \ref{pf}, if the resampling step is omitted, then the weights are replaced by $w^{i}_t= w^{i}_{t-1} g(y_t |x_t^{i},\theta)$.  Particle filtering can also be used for Bayesian parameter inference since it provides an unbiased estimate of the incremental likelihood: 
\begin{align}\label{inclik}
   p(y_{t}|y_{1:t-1},\theta)\approx \hat{ p}_{N_x}(y_{t}|y_{1:t-1},\theta)=\frac{1}{N_x} \sum_{i=1}^{N_x} w_t^{i}
\end{align}
Using the law of total probability, an SMC estimate of the marginal likelihood at $\theta$ follows:
\begin{align}\label{malik}
   p(y_{1: t}|\theta)\approx \hat{ p}_{N_x}(y_{1: t}|\theta)=\prod_{s=1}^t\hat{ p}_{N_x}(y_{s}|y_{1:s-1},\theta)=\prod_{s=1}^t \left(\frac{1}{N_x} \sum_{i=1}^{N_x} w_s^{i} \right).
\end{align}

\subsubsection{\texorpdfstring{SMC$^2$ scheme}{SMC2 scheme}}

Given a weighted sample $\{\theta^m_t, \omega^m_t \}_{m=1}^{N_{\theta}}$ from $p(\theta \vert  y_{1:t})$, the SMC$^2$ algorithm reweights each $\theta$-particle using  an unbiased estimate of the incremental likelihood $p(y_t \vert  y_{1:t-1}, \theta^m_t)$, obtained from the Bootstrap particle filter. Alternative estimates from other filters, such as the auxiliary particle filter, can also be employed to improve efficiency~\citep{Golightly2018}. As we accrue observations through time, there is a risk that only a small number of $\theta$-particles will hold a significant proportion of the weight. To address this, a resampling step is performed to generate a new set of particles from the dominant ones, followed by a PMCMC move step, such as the particle Marginal Metropolis-Hastings (PMMH) kernel, which preserves the target posterior distribution \citep{Andrieu2010}.  The combination of the resampling and move steps is referred to as the ``rejuvenation'' step. In this study, new parameters are proposed from a normal distribution $\theta^* \sim q(\theta^*|\theta^m_t):= \mathcal{N}(\hat{\mu}_t, c\hat{\Sigma}_t)$, where $c$ is a scaling factor. The mean and variance-covariance matrix are given by:
\begin{align}
 \hat{\mu}_t = \dfrac{1}{\sum_{m=1}^{N_{\theta}} \omega^m_t} \sum_{m=1}^{N_{\theta}} \omega^m_t \theta^m_t, \text{ and }
 \hat{\Sigma}_t = \dfrac{1}{\sum_{m=1}^{N_{\theta}} \omega^m_t} \sum_{m=1}^{N_{\theta}} \omega^m_t (\theta^m_t- \hat{\mu}_t)(\theta^m_t - \hat{\mu}_t)^\top,
\end{align}
This allows us to draw new parameter particles in the region where most of the probability mass is located \citep{chopin2002sequential}. 

The $\mathrm{SMC}^2$ scheme is outlined in Algorithm \ref{SMC2}, and the model evidence can be estimated using the following expression:
\begin{align}
    \hat{Z}_t =\prod_{s=1}^{t}\left(\dfrac{1}{\sum_{m=1}^{N_{\theta}} \omega^m_{s-1}} \sum_{m=1}^{N_{\theta}} \omega^m_{s-1} \hat{p}_{N_x}(y_s | y_{1:s-1}, \theta^m_s)\right),
\end{align}
where $ \hat{p}_{N_x}(y_s | y_{1:s-1}, \theta^m_s)$ is the likelihood computed using the particle filter, as described in equation  \eqref{inclik}. Each rejuvenation (step \ref{pmcmc}-\ref{move} of the Algorithm \ref{SMC2}) performs $M$ successive PMMH moves on each particle $\theta^{m}_t$. The resampling step is similar to the one used in Algorithm \ref{pf} but is performed only when a degeneracy criterion is met.

A common way is to estimate the degree of degeneracy by computing the effective sample size (ESS) :
\begin{align}
    \text{ESS}_{\theta}=\left(\sum_{m=1}^{N_{\theta}}\omega^m_t\right)^2 \bigg/  \sum_{m=1}^{N_{\theta}}(\omega^m_t)^2.
\end{align}
The rejuvenation step is triggered if the ESS$_{\theta}$ falls below a specified threshold, typically if ESS$_{\theta} < \tau_R N_{\theta}$, where $\tau_R \in (0, 1)$ is a user-defined parameter. The choice of $\tau_R$ controls the frequency of rejuvenation steps: a higher threshold leads to more frequent rejuvenation, increasing computational cost, while a lower threshold reduces the frequency, potentially allowing particle degeneracy to persist. In practice, an ESS threshold of 50\% ($\tau_R = 0.5$) is commonly used~\citep{Chopin2013,Jacob2015}.

\begin{algorithm}[H]
\caption{SMC$^2$ }\label{SMC2}
Operations involving index $m$ must be performed for $m = 1,..., N_{\theta}$.

The indices $a_{1:N_{\theta}}$ define the ancestral particles after the resampling.

\textbf{Inputs:} Observation: $y_{1:T}$, Prior: $p(\theta)$, Number of state particles: $N_x$, Number of parameter particles: $N_{\theta}$, Resample threshold : $\tau_R$, Move step: $M$

\textbf{Output:}   Particle set: $\left\{\theta^m_{0:t}, \omega^m_{0:t}, x_{0:t, \theta^m}^{1: N_x}, w_{0:t, \theta^m}^{1: N_x}\right\}_{m=1}^{N_{\theta}}$.  

\hrulefill
\begin{algorithmic}[1]
    \State Sample a initial particles $\theta^m_0 \sim p(\theta)$ and set $\omega^m_0=1/N_{\theta}$.
    \For{$t=1$ to $T$}
        \State  Perform iteration $t$ of the BPF to obtain $\left\{x_{t, \theta^m_t}^{1: N_x}, w_{t, \theta^m_t}^{1: N_x}\right\}$ and $\hat{p}_{N_x}(y_t | y_{1:t-1}, \theta^m_t)$. 
        \State Update importance weights: $\omega^m_t=\omega^m_{t-1} \hat{p}_{N_x}(y_t | y_{1:t-1}, \theta^m_t), \quad \omega^m_t=\omega^m_t/\sum_{j=1}^{N_{\theta}} \omega^j_t$
        \If{ESS$_{\theta} <\tau_R N_{\theta}$}
            \State Sample new indices $a_{1:N_{\theta}} \sim \text{Resample}(\omega^{1: N_{\theta}}_t)$
            \State Set $\left\{\theta^m_t, \omega^m_{t}\right\}:=\left\{\theta_t^{a_m}, 1/N_{\theta}\right\}$, $\left\{x_{t, \theta^m_t}^{1: N_x}, w_{t, \theta^m_t}^{1: N_x}\right\}:=\left\{x_{t, \theta_t^{a_m}}^{1: N_x}, w_{t, \theta^{a_m}_t}^{1: N_x}\right\}$ 
            \State\label{pmcmc} Propose $\theta^* \sim q(\theta^*|\theta^m_t)$ and perform BPF with $N_x$ particles given  $\theta^*$ to obtain $ \hat{p}_{N_x}(y_{1: t} | \theta^*)$ \\
            \State Compute : 
              $\alpha=\min \left\{1, \dfrac{ \hat{p}_{N_x}(y_{1: t} | \theta^*)p(\theta^*)}{\hat{p}_{N_x}(y_{1: t} | \theta^m_t)p(\theta^m_t)}\times \dfrac{q(\theta^m_t|\theta^*)}{q(\theta^*|\theta^m_t)} \right\}$\\
            \State\label{move}  Set $\big(\theta_t^m, x_{t, \theta^m_t}^{1: N_x}, w_{t, \theta^m_t}^{1: N_x}\big)=\begin{cases}
               \big(\theta^*, x_{t, \theta^*}^{1: N_x}, w_{t, \theta^*}^{1: N_x}\big) ~\text{ with probability } ~\alpha\\
            \big(\theta_t^m, x_{t, \theta^m_t}^{1: N_x}, w_{t, \theta^m_t}^{1: N_x}\big) ~\text{ with probability }~1-\alpha
            \end{cases}$
        \EndIf
     \EndFor
\end{algorithmic}
\end{algorithm}

\subsection{\texorpdfstring{Online SMC$^2$}{Online SMC2}}

While SMC$^2$ enables sequential estimation of parameters as new observations arrive, it is not a truly online method. The PMMH step, which has a computational complexity of $\mathcal{O}(tN_{\theta}N_x)$, requires calculating the marginal likelihood $p(y_{1:t}|\theta)$. Consequently, the computational cost increases with the number of observations $t$, making the method increasingly expensive as more data accumulates. Although  \cite{Chopin2013} demonstrates that, in theory, the frequency at which the rejuvenation step is performed will decrease over time, the algorithm may still be impractical for some real-time applications. \cite{vieira2018bayesian} proposes to implement SMC$^2$ in an online manner (O-SMC$^2$) by calculating the likelihood using only a fixed-size window of past observations. This restriction keeps the computational cost of a likelihood calculation constant. By considering a fixed window of observations of size $t_k > 0$ in the PMMH, we can ensure that the online part of SMC$^2$ will not be altered, and the online approximation will target the posterior of the parameters given the observation window $p(\theta | y_{t-t_k+1:t})$. This reduces the computational cost of each PMMH iteration from  $\mathcal{O}(tN_{\theta}N_x)$ to $\mathcal{O}(t_kN_{\theta}N_x)$.

Hence, in step \ref{pmcmc} of Algorithm \ref{SMC2}, once we have reached $t > t_k$, the Metropolis-Hastings acceptance probability will now take the form:
\begin{align}
    \min \left\{1, \dfrac{\hat{p}_{N_x}(y_{t-t_k+1:t} | \theta^*) p(\theta^*)}{\hat{p}_{N_x}(y_{t-t_k+1:t} | \theta^m_t) p(\theta^m_t)} \times \dfrac{q(\theta^m_t|\theta^*)}{q(\theta^*|\theta^m_t)} \right\},
\end{align}
where the estimate of the marginal likelihood within the time window is computed similarly to \eqref{malik}, but this time using the restricted particle set $x_{t-t_k+1:t}^{1:N_x}$ (see \ref{AppB} for further details) :
\begin{align}
    \hat{p}_{N_x}(y_{t-t_k+1:t} | \theta) = \prod_{s=t-t_k+1}^t \left(\frac{1}{N_x} \sum_{i=1}^{N_x} w_s^{i} (x_{s-1}^{i}, x_s^{i}) \right).
\end{align}
This restriction is possible under the assumption that static parameters will remain constant indefinitely \citep{vieira2018bayesian}. The O-SMC$^2$ scheme appears to be particularly amenable to parallelization over parameter particles, as the rejuvenation step can be performed separately for each parameter value.

\begin{figure}[H]
    \centering
    \includegraphics[scale=1]{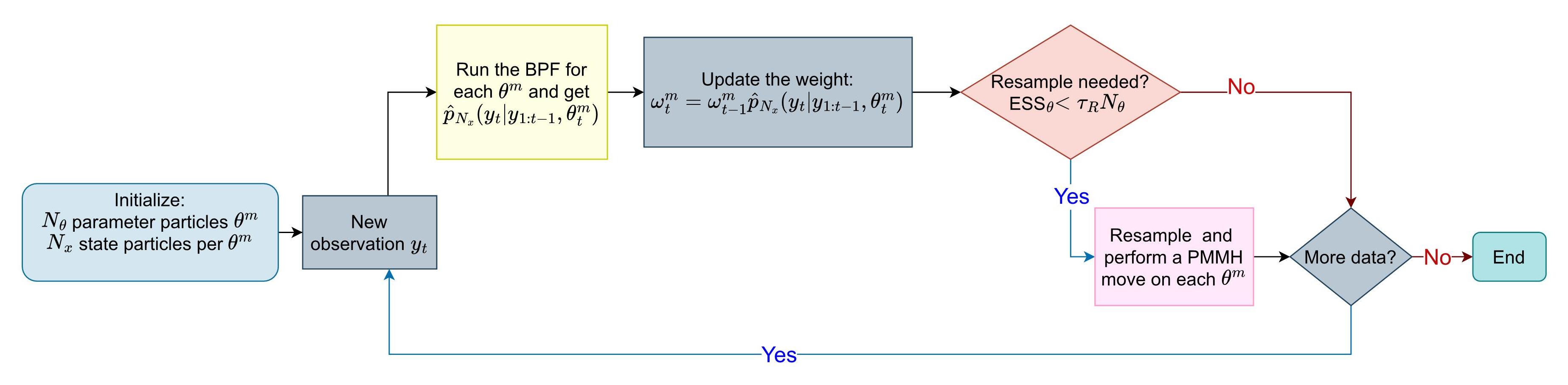}    \caption{\footnotesize Flowchart of the SMC$^2$/O-SMC$^2$ algorithm. }
    \label{figure:Fig3}
\end{figure}
Figure~\ref{figure:Fig3} illustrates the flow of the SMC$^2$/O-SMC$^2$ algorithm. At each time step, a new observation $y_t$ arrives. A particle filter is run for each parameter particle $\theta^{m}$ to estimate the likelihood and update the weights. If resampling is needed, it is followed by a PMMH rejuvenation step. The procedure repeats as new observations arrive, allowing for sequential joint inference of parameters and latent states.

\subsection{Structural identifiability}

Prior to presenting the inference results using the O-SMC$^2$ methodology in Section~\ref{sec3}, we conduct a structural identifiability analysis of the SEIR model parameters.  This analysis employed a deterministic mean-field approximation \citep{roberts2015nine} of the stochastic model given in Equation~\eqref{stoseir} and was performed using the \texttt{StructuralIdentifiability.jl} package in Julia \citep{dong2023differential,liyanage2025tutorial}. The results indicate that the parameters $\beta$, $\sigma$, and $\gamma$ are globally identifiable from ideal observations of the daily incidence. This provides a theoretical basis supporting the feasibility of parameter estimation under perfect data conditions. Further details of the model formulation and the output function used are provided in Appendix \ref{AppA}.

However, it is important to emphasize some limitations inherent in structural identifiability analysis. The analysis makes the assumptions of noise-free, continuously reported and complete observations. These assumptions will rarely hold in practical epidemiological settings. Epidemiological data are often sparse, noisy, and collected at discrete time points (for example, daily or weekly reports), and may have missing information. These challenges are especially pronounced in the early stages of an outbreak.

In the context of sequential parameter estimation, when only a small number of observations have been accrued at the early stages of an outbreak, we expect to see high uncertainty, wide credible intervals, and possible parameter confounding \citep{sauer2021identifiability}. Thankfully, this is usually observed only during early phases. With the incorporation of informative priors, these challenges are somewhat mitigated and due to the adaptive qualities of sequential inference, they gradually resolve as we accrue more data. We may also experience challenges during periods of abrupt change in disease dynamics. Parameters may become highly correlated and hence poorly identified. This again can be mitigated through the use of informative priors. 

In summary, the conclusions drawn from an identifiability analysis are subject to often unrealisable assumptions in practical settings. In our context, idealized structural analyses should be viewed as necessary but not sufficient conditions for practical identifiability.

\section{Experimental and Real-world data analysis}\label{sec3}
To illustrate the methodology described in the previous sections, we conduct a series of experiments using both simulated and real-world data. The algorithm is implemented in Python version 3.10.12 and was executed on a desktop computer with an Intel Core i7-1300H processor and a 3.40 GHz clock speed. The code is available at \href{https://github.com/Dhorasso/smc2-seir-model}{https://github.com/Dhorasso/smc2-seir-model}.

\subsection{Experimental setup}
We begin by applying the filtering algorithm to two synthetic datasets generated using the stochastic SEIR model introduced in Section \ref{sec3}.  The simulation period spans $T = 100$ days, with a daily time step of $\delta t = 1$ day. The algorithm utilizes $N_{\theta} = 400$ parameter particles, $N_x = 200$ state particles, with $M=5$ successive PMMH moves and a window size of $t_k = 20$ days. In Experiment 1, all model parameters remain constant, while Experiment 2 introduces a time-varying transmission rate. We assume that the data on daily new cases are the realization of a Poisson distribution: $y_t | x_t \sim \mathrm{Poisson}(Y_{EI}(t))$, where  $Y_{EI}(t)$ is the number of individual moving from $E$ to $I$ at time $t$.

\subsubsection{Experiment 1}
In this experiment, we assume constant parameters for the SEIR model: $\beta = 0.6$, $\sigma = \frac{1}{3}$, and $\gamma = \frac{1}{5}$. The total population size is $N = 6000$, with one initially infected individual. The latent state at time $t$ is represented by $x_t = (S_t, E_t, I_t, R_t)$, and the parameters $\theta = (\beta, \sigma, \gamma)$ are to be estimated. The initial particle state vector in the O-SMC$^2$ algorithm is set to $(S_0, E_0, I_0, R_0) = (N - I_0, 0, I_0, 0)$, with $I_0$ drawn from a discrete uniform distribution, $I_0 \sim \mathcal{U}({0, \dots, 5})$. A uniform prior on the interval $[0, 1]$ is assigned to each of the parameters $\beta$, $\sigma$, and $\gamma$.

\begin{figure}[H]
	        \centering
  \includegraphics[scale=0.95]{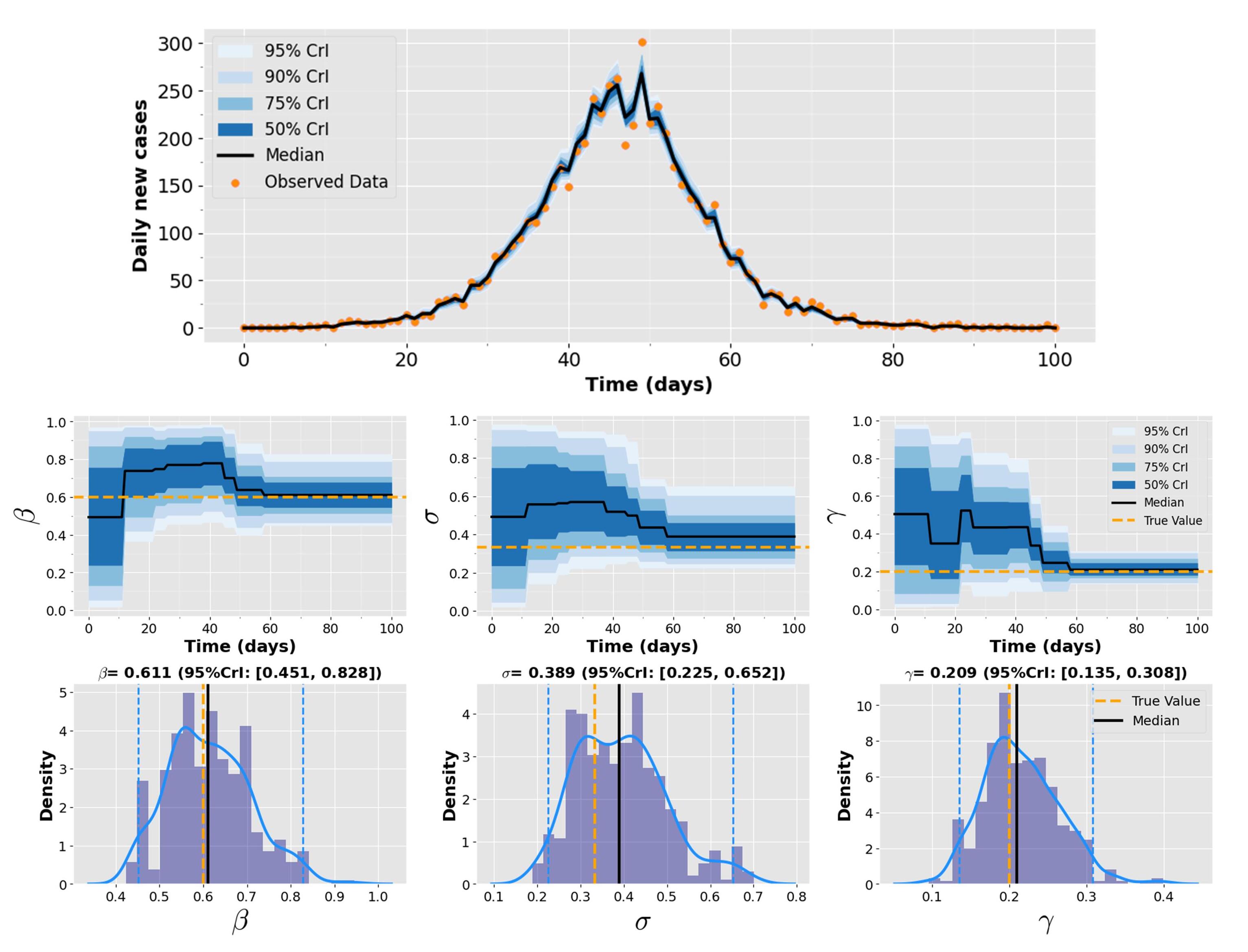}
    \caption{\footnotesize  State and parameter inference for Experiment 1. 
    Top row: Filtering estimates of daily new infections, with observed values shown as orange dots. 
   Middle row: Time series of filtering estimates for parameters $\beta$, $\sigma$, and $\gamma$. 
     Bottom row: Posterior distributions of $\beta$, $\sigma$, and $\gamma$ at the final time step, with 95\% credible intervals in dashed blue lines. In rows 1 and 2, the solid black lines show the median estimate; shaded blue regions indicate posterior uncertainty quantiles. True parameter values are represented by orange dashed lines.}
    \label{figure:Fig4}
\end{figure}

Figure \ref{figure:Fig4} displays the sequential inference of daily new cases of infection together with the static parameters. We can observe that the predicted new cases are in good agreement with the data. The results also indicate that all true parameter values fall within the 95\% credible interval, closely aligning with the posterior median as we evolve over time. Initially, during the epidemic's onset, the data lacks sufficient information within the time interval $[0,20]$, resulting in widespread parameter uncertainty. However, as the epidemic approaches its peak, the inference algorithm captures the dynamics well, notably reducing parameter uncertainty. As the dynamics begin to slow around time $t = 60$, the posterior uncertainty stabilizes.

\subsubsection{Experiment 2}
In this experiment, we introduce variability in the transmission rate to simulate a more realistic scenario. Synthetic data are generated using a time-varying $\beta=\beta(t)$, modeled as:
\begin{align} 
\beta(t) = 0.3 \exp\left(\sin\left(\frac{2 \pi t}{55}\right) - \frac{t}{80}\right), 
\end{align}
while the parameters $\sigma = \frac{1}{2}$ and $\gamma = \frac{1}{7}$ remain constant. The total population size is $N = 200,000$, with five initially infected individuals. This experiment assesses the ability of the O-SMC$^2$ algorithm to capture state and static parameters under high transmission variability.

To accommodate the non-constant transmission rate, we model $\beta(t)$ in the model \eqref{stoseir} as a geometric Brownian motion:
\begin{equation} \label{rw}
    \beta(t) = e^{\log (\beta(t-1)) + \varepsilon_{t}}, \quad \varepsilon_{t} \sim \mathcal{N}(0, \nu_{\beta}^2),
\end{equation}
where $\nu_{\beta}$ governs the variability in the transmission rate. The latent state vector is now $x_{t} = (S_{t}, E_{t}, I_{t}, R_{t}, \beta(t))$, and the parameters to estimate are $\theta = (\sigma, \gamma, \nu_{\beta})$. The O-SMC$^2$ algorithm uses the same initial particle state as in Experiment 1, with $\beta(0) \sim \mathcal{U}(0.2, 0.5)$. 

\begin{figure}[H]
	        \centering
 \includegraphics[scale=0.95]{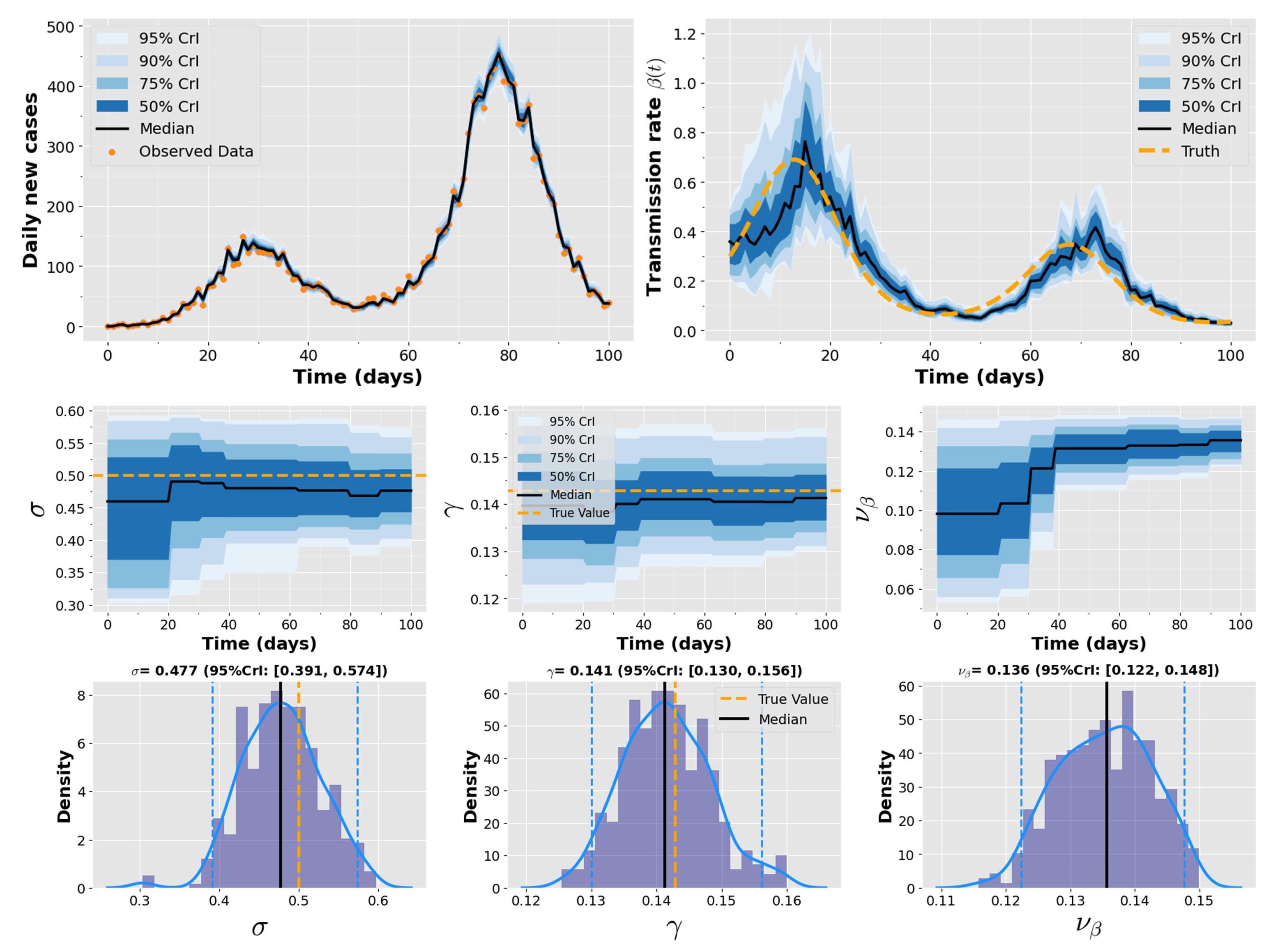}
    \caption{\footnotesize  State and parameter inference for Experiment 2. 
    Top row: Left - Filtering estimates of daily new infections with observed values in orange; Right - Filtering estimate of the time-dependent transmission rate $\beta(t)$. 
   Middle row: Filtering estimates of $\sigma$, $\gamma$, and $\nu_{\beta}$ over time. 
     Bottom row: Posterior distributions of $\sigma$, $\gamma$, and $\nu_{\beta}$  the final time step, with 95\% credible intervals in dashed blue lines. In rows 1 and 2, the solid black lines show the median estimate; shaded blue regions indicate posterior uncertainty quantiles. True parameter values are represented by orange dashed lines.}
    \label{figure:Fig5}
\end{figure}
Priors for the parameters are $\sigma \sim \mathcal{U}(0.3, 0.6)$, $\gamma \sim \mathcal{TN}_{[0, 1]}(0.14, 0.01)$, and $\nu_{\beta} \sim \mathcal{TN}_{[0.05, 0.15]}(0.1, 0.05)$. A more restrictive prior on $\gamma$ is used due to its high correlation with the transmission rate $\beta(t)$, especially during abrupt changes in transmission dynamics. Fluctuations in new cases can reflect changes in $\beta(t)$, but could also be influenced by $\gamma$  as shown in \citep{inouzhe2023}. The prior for $\nu_{\beta}$ allows for rapid changes in the transmission rate, ensuring the model can adapt to sudden shifts in disease dynamics and aligns with values observed in the literature \citep{funk2018realtime, calvetti2021}.

Figure \ref{figure:Fig5} demonstrates the ability of the method to track multiple outbreak waves and capture the trend in a time-varying transmission rate, albeit with a minor lag. This slight delay is expected, as the analysis relies on daily case data (i.e., the number of newly infectious individuals transitioning from $E$ to $I$) rather than exact counts of new infections (i.e., individuals moving from $S$ to $E$), which are often unknown during an epidemic. As with Experiment 1, the true parameter values lie within the 95\% credible interval, indicating reliable parameter estimation even under conditions of variable transmission.

To evaluate the performance of the O-SMC$^2$ algorithm, we tested a range of window sizes $t_k$, keeping the priors and number of particle configurations consistent across experiments. Tables \ref{TableC2} and \ref{TableC3} in Appendix \ref{AppC} show that even with smaller window sizes, the algorithm produces reasonable parameter estimates that are close to the true values, with credible intervals comparable to those of standard SMC$^2$  approach (i.e., when all past observations are used at each time step), although slightly larger for smaller window sizes. Notably, the computational cost decreases substantially with smaller windows, making O-SMC$^2$ well-suited for real-time or resource-limited settings. These findings align with previous research (\cite{vieira2018bayesian}, Chapter 15). Figures \ref{figure:FigC1}-\ref{figure:FigC2} in Appendix \ref{AppC} display the posterior densities of the parameters at the final time step, demonstrating consistency with estimates obtained using SMC$^2$.

\subsection{Real case study: Analysis of COVID-19 in Ireland}

Since the onset of the COVID-19 pandemic in December 2019 in Wuhan, China, the World Health Organization (WHO) has reported over 772 million confirmed cases globally and more than 69 million deaths \citep{who2020covid}. The rapid transmission and widespread nature of the virus led the WHO to declare it a pandemic. In \citet{cazelles2021dynamics}, particle MCMC has been employed to reconstruct the COVID-19 epidemic's trajectory in Ireland. However, it is worth noting that while this method efficiently tracks observed data, it remains an offline approach and may not be suitable for real-time decision-making when rapid responses are required.

\subsubsection{Model}
We present a fully discrete stochastic extended-SEIR model to study the spread of COVID-19,  with a structure designed to capture changes in transmission dynamics during the pandemic. In this model, individuals transition through compartments representing different stages of infection: $S_{t}$ representing susceptible individuals, $E_{t}$ for those who are exposed and infected but not yet infectious, $A_{t}$ for asymptomatic infectious individuals, $I_{t}$ for symptomatic infectious individuals, and finally, $R_{t}$ is the number of individuals recovering or succumbing to the virus. The following discrete equations describe the dynamics of the model:
\begin{align} \label{covm}
    S_{t+\delta t} &= S_{t} - Y_{SE}(t), \notag \\
    E_{t+\delta t} &= E_{t} + Y_{SE}(t) - Y_{E}(t), \notag \\
    A_{t+\delta t} &= A_{t} + Y_{EA}(t) - Y_{AR}(t), \notag \\
    I_{t+\delta t} &= I_{t} + Y_{E}(t) - Y_{EA}(t) - Y_{IR}(t), \\
    R_{t+\delta t} &= R_{t} + Y_{AR}(t) + Y_{IR}(t), \notag
\end{align}
where $Y_{AB}(t)$  represents the number of individuals moving from compartment $A$ to compartment $B$ at time $t$ and these quantities are defined through the binomial random variables:
\begin{align} \label{covb}
&Y_{SE}(t) \sim \mathrm{Bin}\left(S_{t}, 1 - \exp\left(-\beta(t) \frac{(I_{t} + r_{A} A_{t})}{N}  \delta t\right)\right), \notag \\
&Y_{E}(t) \sim \mathrm{Bin}\left(E_{t}, 1 - e^{-\sigma  \delta t}\right), \quad
Y_{EA}(t) \sim \mathrm{Bin}\left(Y_{E}(t), p_{A}\right), \\
& Y_{AR}(t) \sim \mathrm{Bin}\left(A_{t}, 1 - e^{-\gamma  \delta t}\right), \quad Y_{IR}(t) \sim \mathrm{Bin}\left(I_{t}, 1 - e^{-\gamma  \delta t}\right).\notag
\end{align}
The model assumes a homogeneously mixing population of constant size:
\begin{equation}
N = S_{t} + E_{t} + A_{t} + I_{t} + R_{t}.
\end{equation}

The models presented in \eqref{covm}-\eqref{covb}, do not explicitly incorporate vaccination. This is because they are primarily designed to focus on the spread of the virus and the impact of various non-pharmaceutical interventions during a period when vaccines are either unavailable or not yet widely deployed.  However, the effect of vaccination can easily be introduced by adding a compartment for vaccinated individuals and adjusting the transmission dynamics to reflect partial or complete immunity. Allowing the transmission rate to be time-varying seems reasonable since, throughout the epidemic, the intensity of disease transmission will tend to vary due to individuals' behavioral changes or government lockdown measures. Hence, we consider a geometric random walk (RW) model for the transmission rate:
\begin{equation} \label{rwm} 
    \beta(t) = e^{\log (\beta(t-1)) + w_{t}}, \quad w_{t} \sim \mathcal{N}(0, \nu_{\beta}^2),
\end{equation}
where $\nu_{\beta}$ is the parameter controlling the innovation in the transmission rate, consistent with previous study of COVID-19 in Ireland  \citep{cazelles2021dynamics}. Giving the distribution of $\beta(t)$, we can sequentially estimate the distribution of the effective reproduction number ($R_{\text{eff}}(t)$) that refers to the average number of secondary infections caused by a single infected individual during their infectious period in a population where some individuals are immune or other interventions have been implemented:
\begin{align}
R_{\text{eff}}(t) = \beta(t) \left( \frac{(1 - p_{A}) + p_{A}  r_{A}}{\gamma} \right) \frac{S_t}{N}.
\end{align}
In general, when $R_{\text{eff}}(t) > 1$, the number of observed cases will increase, while it will decrease if $R_{\text{eff}}(t) < 1$. Consequently, policymakers can determine whether to relax or strengthen control measures based on whether $R_{\text{eff}}$ falls below the self-sustaining threshold of $1$.

\subsubsection{Data and model fitting}
The data we use are daily confirmed case counts extracted from an up-to-date database of the Irish Health Protection Surveillance Centre (HPSC) (\href{https://COVID19ireland-geohive.hub.arcgis.com/}{https://COVID19ireland-geohive.hub.arcgis.com/}). To integrate these observations into our model, we define the variable: $C^I_{t}$ representing the cumulative number of infectious cases. This variable evolves according to the equation:
\begin{equation}
    C^I_{t+1} = C^I_{t} + Y_{E}(t) - Y_{EA}(t).
\end{equation}
Hence, $\mu_{t} = C^I_{t+1} - C^I_{t}$ represents the daily expected number of cases. Since daily reported case counts are subject to observation error and typically exhibit overdispersion (where the conditional variance exceeds the conditional mean), we model the observed daily cases, $Z^I_t$, using a normal approximation of the negative binomial distribution (chosen for computational
efﬁciency, \cite{funk2018realtime}). The negative binomial distribution is a generalization of the Poisson distribution that can account for overdispersion in count data. Specifically, for the negative binomial distribution, the mean is $\mu_t$ and the variance is $\mu_t(1 + \phi \mu_t)$, where $\phi$ is an overdispersion parameter. To avoid singularities when $\mu_{t} = 0$, we followed the approach of  \cite{funk2018realtime},  rounding variances smaller than 1 up to 1.  The observed daily case counts follow a negative binomial distribution: 
\begin{equation}\label{ov}
Z^I_t\sim\mathcal{N}(\mu_{t},\mu_{t}(1+\phi \mu_{t})). 
\end{equation}
We estimate the spread of COVID-19 in Ireland from February 29th, 2020, to February 28th, 2021, using the daily new case count data over 365 days. The Irish population in 2020 was estimated to be $N= 4,965,439$.  The O-SMC$^2$ uses $N_{\theta} = 1000$ parameter particles, $N_x = 500$ state particles, $M = 5$ successive PMMH moves and a window size of $t_k = 80$ days. Table \ref{TableD1} in Appendix \ref{AppD} gives the description and prior of the inferred parameters used in the model \eqref{covb}.  These priors are based on previous studies of COVID-19 dynamics in Ireland \citep{IEMAG2020, cazelles2021dynamics}. For the choice of prior distributions for the hyperparameters, $\nu_{\beta}$ (which governs the variability in the transmission rate) and $\phi$ (the overdispersion parameter), we adopt settings that allow for rapid changes in the transmission rate, ensuring the model can adapt to sudden shifts in disease dynamics, as discussed in \citep{funk2018realtime}. Regarding initial states, we set: $ S_{0} =N- E_{0}-A_{0}$, $E_{0}=1$, $A_{0}\sim\mathcal{U}(\{10,\dots,50\})$ and $ I_{0}= R_{0}=0$. The execution time for the 365 days considered here was less than 5 hours, making it suitable for practical real-time applications.

\begin{figure}[H]
	        \centering
 \includegraphics[scale=1]{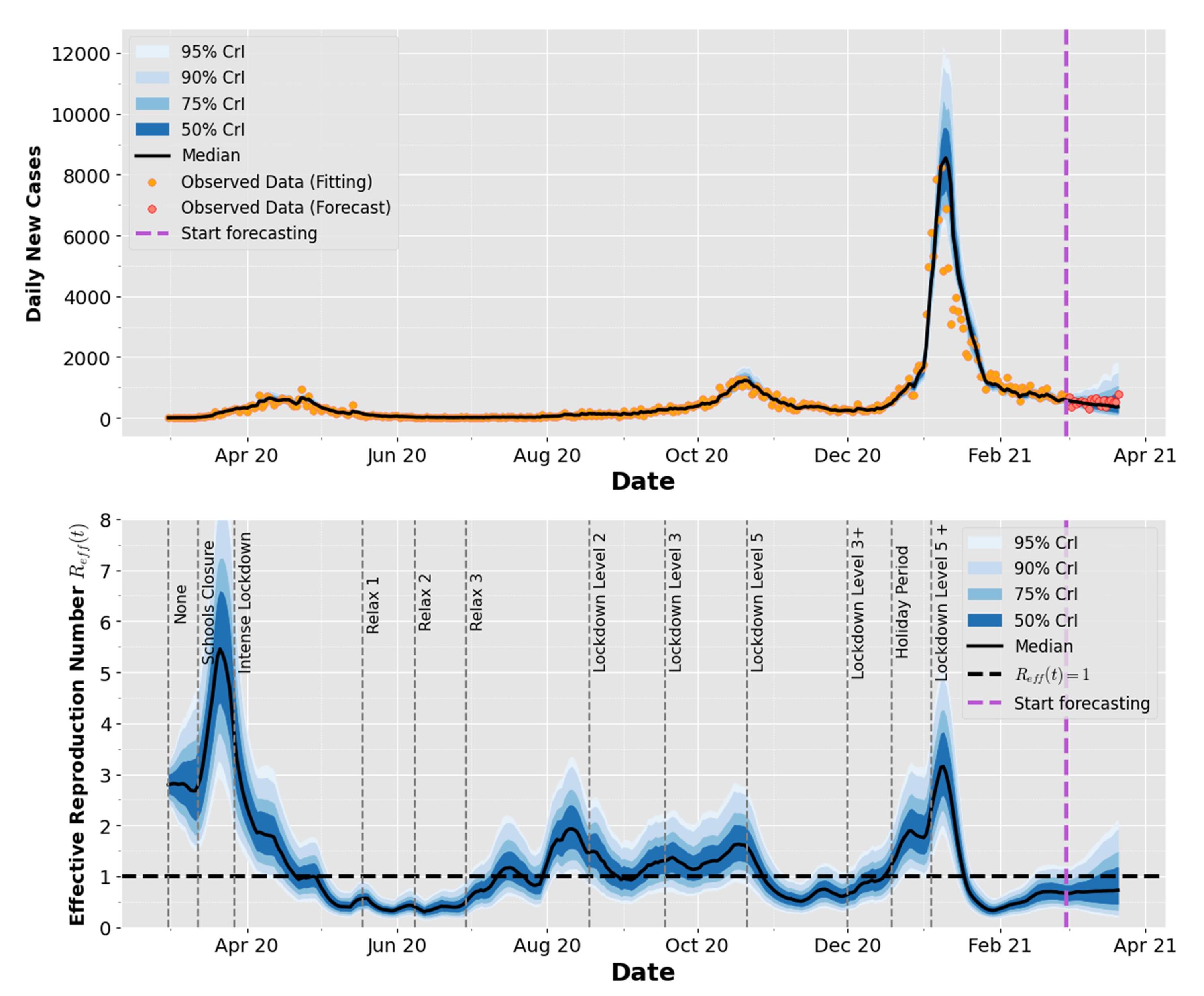}
	\caption{\footnotesize  Model fitting with COVID-19 data in Ireland. The top row shows the estimated daily number of new cases. The observed data used for model fitting are shown as orange dots, while the red dots represent observed data that were excluded from fitting and used solely for forecast validation. The bottom row presents the weekly average of the time-dependent effective reproduction number. The solid black lines show the median estimate; shaded blue regions indicate posterior uncertainty quantiles.  The vertical grey dashed lines represent the different government interventions. The vertical dashed violet line marks the last date of observations, beyond which the model is used to forecast the next 21 days.}
\label{figure:Fig6}
\end{figure}

Figure \ref{figure:Fig6} illustrates the model's fit to the daily number of COVID-19 cases reported in Ireland. The top plot shows the daily new cases, with three noticeable peaks: one in mid-April 2020, another in mid-November 2020, and a larger peak in early January 2021. The model successfully captures these three waves, with observed data (orange dots) falling within the 95\% credible interval on most days. The fitting period achieved a mean absolute error (MAE) of 139.59 and a continuous ranked probability score (CRPS) of 105.78. These evaluation metrics are defined in detail in Appendix \ref{AppD}. However, a sudden drop in reported cases around mid-January 2021 is not fully captured, likely due to real-world factors such as reporting delays or testing changes not included in the model. Beyond February 28th, 2021, the model projects cases forward for 21 days, excluding data from April 1 (red dots) from the fitting process. The forecast period shows good predictive accuracy with an MAE of 113.71 and a CRPS of 98.45, and the observed data fall within the 95\% credible interval, indicating strong agreement between predictions and actual trends.

The bottom plot in Figure \ref{figure:Fig6} shows the weekly-averaged effective reproduction number, $R_{\text{eff}}(t) $, with corresponding daily estimates (not averaged) presented in Figure \ref{figure:FigD1} in Appendix \ref{AppD}. As anticipated, $R_{\text{eff}}(t) $ fluctuates in response to changes in the epidemic dynamics. The impact of the first lockdown is clearly evident, as it drives $R_{\text{eff}}(t) $ below one, effectively ending the first wave. A gradual increase in $R_{\text{eff}}(t) $ occurred in June 2020, due to the easing of restrictions. $R_{\text{eff}}(t) $ remained above one until mid-October, corresponding to the second wave of the epidemic. The second round of lockdown measures caused $R_{\text{eff}}(t) $ to decline by November 2020, falling below one once more. The emergence of the more transmissible Alpha variant, combined with increased social interactions during the holiday season, resulted in a sharp rise in cases toward the end of 2020. This was followed by a third round of lockdown measures, which once again brought $R_{\text{eff}}(t) $ below one (see the supplementary material of \cite{jaouimaa2021age} for a description of the individual lockdown measures).

\begin{figure}[H]
	        \centering
           \includegraphics[scale=1]{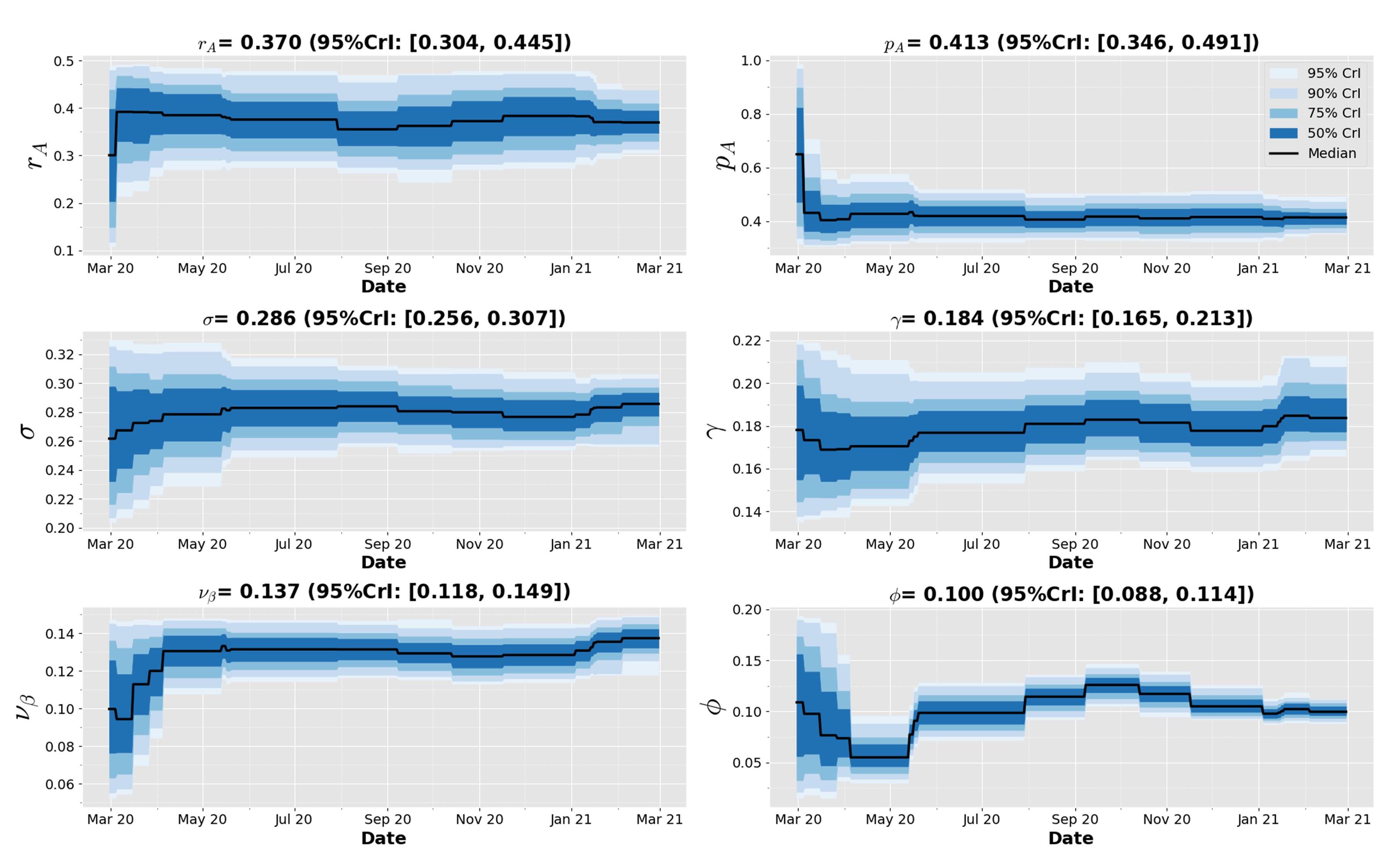}
	\caption{\footnotesize  Filtered estimate of the parameter in the COVID-19 model using O-SMC$^2$ algorithm. The solid black lines represent the median estimate; shaded blue regions indicate posterior uncertainty quantiles.}
 \label{figure:Fig7}
\end{figure}

In Figure \ref{figure:Fig7}, we present the sequential estimation of the parameters of model \eqref{covm} over 365 days of the pandemic. The graphs display the posterior medians of the parameters, accompanied by their respective uncertainty intervals at each time step. The posterior distributions of the parameters obtained at the final time step (see Table \ref{TableD1} in Appendix \ref{AppD}) are aligned with the results found in \citet{cazelles2021dynamics} using PMCMC methods.

Figure \ref{figure:FigE1} in Appendix \ref{AppE} illustrates the application of the O-SMC$^2$ for real-time monitoring, using partial datasets spanning four distinct time periods:  February 29, 2020, to May 30, 2020; May 30, 2020, to August 29, 2020; August 29, 2020, to November 28, 2020; and November 28, 2020, to February 27, 2021. At each stage, the algorithm utilizes the results from the previous O-SMC$^2$ run as prior information to compute the joint posterior $p(x_{i:j}, \theta \mid y_{i:j})$. This sequential approach enables the efficient integration of new data while maintaining computational efficiency. Moreover,  it significantly improves the methodology described in \citet{Dureau2013}, which relied on PMCMC and required the model to be rerun from scratch for each new batch of data.

\section{Conclusion}\label{sec4}
In this paper, we explored the application of an online variant of the Sequential Monte Carlo Squared (O-SMC$^2$) method for real-time inference in stochastic epidemic models. This approach differs from standard SMC$^2$ by integrating a mechanism that approximates the posterior distribution using a fixed window of recent observations, thereby reducing computational demands while maintaining parameter estimation accuracy. This adaptation makes O-SMC$^2$ particularly suitable for dynamic disease modeling, where timely responses are critical. We demonstrated the effectiveness of O-SMC$^2$ in the stochastic SEIR compartment model, validating its performance on both synthetic simulations and real-world COVID-19 data from Ireland. Our results show that O-SMC$^2$, even with a modest window size, effectively tracks epidemic progression and generates robust online estimates of both static and dynamic epidemiological parameters. Moreover, the sequential nature of this method enables the efficient updating of estimates by using results from previous runs as prior information, eliminating the need to rerun the model from scratch as new data becomes available incrementally. This feature underscores the practicality of O-SMC$^2$ for real-time epidemiological surveillance, informing public health authorities on adaptation of interventions based on real-time information assimilation. However, we also acknowledge certain challenges, particularly in the early stages of an outbreak when available data may be sparse and less informative.  During this phase, prior uncertainty, potential parameter non-identifiability, and low case counts can collectively reduce the reliability of estimates. These issues can result in wide credible intervals and greater variability in posterior distributions. One practical consideration is the choice of the fixed window size, $t_k$, which impacts the balance between computational efficiency and the informativeness of the data used for inference. In our experiments, setting $t_k$ to approximately 20\% of the total time series length provided a reasonable compromise.  However, in truly online settings, where data arrives as a continuous stream without a predetermined end, a fixed window size may not be optimal. An adaptive strategy that dynamically adjusts $t_k$, for example,  by increasing the window size when the PMMH acceptance rate fall below a certain threshold, could improve the robustness during periods of rapid change or less informative observations. Exploring such adaptive mechanisms and integrating more flexible modeling approaches are key directions for enhancing the responsiveness and reliability of early-stage inference in real-time epidemic monitoring.

\section*{Acknowledgements}
This publication has emanated from research conducted with the financial support of Taighde Éireann – Research Ireland under Grant number 21/FFP-P/10123. We wish to thank the editor and the anonymous reviewers for their constructive comments and suggestions which greatly improved the quality of this paper.

\appendix

\section{Structural identifiability analysis}\label{AppA}

To assess structural identifiability, we first derived a deterministic mean-field approximation of the stochastic SEIR model. This step is necessary because most identifiability analysis tools require a system of ordinary differential equations (ODEs) as input. Starting from the discrete-time stochastic model, we replaced each binomial transition with its expected value and applied the first-order approximation  $1 - e^{-x} \approx x$ for  $x\ll1$. Taking the limit as $\delta t \to 0,$ , we obtained the following system:
\begin{align}
\frac{dS}{dt} = -\beta \frac{S I}{N},\quad
\frac{dE}{dt} = \beta \frac{S I}{N} - \sigma E, \quad
\frac{dI}{dt} = \sigma E - \gamma I, \quad
\frac{dR}{dt} = \gamma I.
\end{align}
We assumed the observable is  $y_1(t) = \sigma E(t)$, representing daily new infections. All parameters are also assumed to be static throughout the analysis, consistent with standard structural identifiability frameworks.

Structural identifiability refers to the theoretical ability to uniquely recover model parameters from perfect data, given the model structure and observables. This was assessed using the Julia package \texttt{StructuralIdentifiability.jl} \citep{dong2023differential}. The total population size $N$ and the initial conditions were assumed to be known during the analysis. Results (Table~\ref{TableA1}) show that for both the standard SEIR and the SEAIR model, key parameters such as $\beta$, $\sigma$, $\gamma$ and additional parameters in the extended model, are structurally globally identifiable. For the SEIR model, we also found that all parameters remain globally identifiable even when the initial conditions are unknown, provided that N is known. These findings confirm that the selected observables are theoretically sufficient to uniquely identify the parameters prior to applying the O-SMC$^2$ algorithm. For a tutorial introduction to \texttt{StructuralIdentifiability.jl} in epidemic modeling, see \citep{liyanage2025tutorial}.

\begin{table}[H]
\centering
\tiny
\caption{Structural identifiability results.}
\begin{adjustbox}{width=\textwidth}
\begin{tabular}{p{0.48\textwidth} | p{0.48\textwidth}}
\hline
\textbf{SEIR} & \textbf{SEIAR} \\
\hline
\hline
\begin{verbatim}
julia> ode = @ODEmodel(
       S'(t) = - beta*S(t)*I(t)/N,
       E'(t) = beta*S(t)*I(t)/N - sigma*E(t),
       I'(t) = sigma*E(t) - gamma*I(t),
       R'(t) = gamma*I(t),
       y1(t) = sigma*E(t),
       y2(t) = N
   )
julia> assess_identifiability(ode, known_ic = [S,E,I,R])
  S(0)  => :globally
  E(0)  => :globally
  I(0)  => :globally
  R(0)  => :globally
  beta  => :globally
  gamma => :globally
  sigma => :globally
\end{verbatim}
&
\begin{verbatim}
julia> ode = @ODEmodel(
       S'(t) = - beta*S(t)*(I(t)+ra*A(t))/N,
       E'(t) = beta*S(t)*(I(t)+ra*A(t))/N - sigma*E(t),
       A'(t) = pa*sigma*E(t) - gamma*A(t),
       I'(t) = (1-pa)*sigma*E(t) - gamma*I(t),
       R'(t) = gamma*A(t) + gamma*I(t),
       y1(t) = (1-pa)*sigma*E(t),
       y2(t) = N
   )
julia> assess_identifiability(ode, known_ic = [S,E,A,I,R])
  S(0)  => :globally
  E(0)  => :globally
  A(0)  => :globally
  I(0)  => :globally
  R(0)  => :globally
  N     => :globally
  beta  => :globally
  gamma => :globally
  pa    => :globally
  ra    => :globally
  sigma => :globally
\end{verbatim}
\\
\hline
\end{tabular}
\end{adjustbox}
\label{TableA1}
\end{table}

\section{Approximation of the acceptance probability}\label{AppB}

During the rejuvenation step in the algorithm \ref{SMC2}, when a new parameter $ \theta^{*} $ is proposed at time $ t $, an unbiased estimate of the marginal likelihood is computed by running Algorithm \ref{pf} over observations from time 1 to $ t $. However, as time progresses, the computational cost of this procedure grows, making it less feasible for long-time series data. This limitation can be alleviated by considering only the most recent observations \citep{vieira2018bayesian}.

Let the window size be denoted by $ t_k > 0 $. By the law of total probability, the marginal likelihood at time $ t $ can be factorized as:
\begin{align}
    p(y_{1:t} | \theta) &= \left( \prod_{s=1}^{t-t_k} p(y_s | y_{1:s-1}, \theta) \right) \times \left( \prod_{s=t-t_k+1}^t p(y_s | y_{1:s-1}, \theta) \right) \notag \\
    &= p(y_{1:t-t_k} | \theta) \times \left( \prod_{s=t-t_k+1}^t p(y_s | y_{1:s-1}, \theta) \right).
\end{align}
For $s \geq t - t_k$, if we assume that for a sufficiently large $t_k$, the older observations $y_{1:t-t_k}$ have minimal impact on the filtering estimate of the latent state at time $s-1$, i.e., $p(x_{s-1} | y_{1: s-1}, \theta) \approx p(x_{s-1} | y_{t-t_k+1: s-1}, \theta)$, then, from the result in \citep{kantas2015particle}, we have:
\begin{align}
    p(y_s | y_{1:s-1}, \theta) &= \int g(y_s | x_s, \theta) f(x_s | x_{s-1}, \theta) p(x_{s-1} | y_{1: s-1}, \theta) \, dx_{s-1:s} \notag\\
    &\approx \int g(y_s | x_s, \theta) f(x_s | x_{s-1}, \theta) p(x_{s-1} | y_{t-t_k+1: s-1}, \theta) \, dx_{s-1:s} \notag\\
    &= p(y_s | y_{t-t_k+1:s-1}, \theta ).
\end{align}
This approximation yields to:
\begin{align}
    p(y_{1:t} | \theta) \approx p(y_{1:t-t_k} | \theta) \times p(y_{t-t_k+1:t} | \theta),
\end{align}
with:
\begin{align}
  p(y_{t-t_k+1:t} | \theta) = \prod_{s=t-t_k+1}^t p(y_s | y_{t-t_k+1:s-1}, \theta) \approx \prod_{s=t-t_k+1}^t p(y_s | y_{1:s-1}, \theta).
\end{align}
Given the current $\theta$-particle, the new parameter $\theta^{*}$ at time $t$ is drawn from the region where most of the probability mass is located \citep{chopin2002sequential}. Under the assumption that the target parameter vector remains constant over time, and provided that the window size $t_k$ is large enough to capture the necessary temporal variability in the observations $y_{t-t_k+1:t}$, it is reasonable to assume that the historical likelihoods  $p(y_{1:t-t_k} | \theta)$ and $p(y_{1:t-t_k} | \theta^{*})$ are approximately equal. Consequently, their contribution to the likelihood ratio becomes negligible. Under this assumption, the full likelihood ratio simplifies to:
\begin{align}
    \dfrac{p(y_{1:t} | \theta^{*})}{p(y_{1:t} | \theta)} \approx \dfrac{p(y_{t-t_k+1:t} | \theta^{*})}{p(y_{t-t_k+1:t} | \theta)}.
\end{align}
Thus, using the SMC estimate of the incremental likelihood for a fixed number of particles $N_x$ (see Equation \eqref{inclik}),
we estimate the windowed likelihood $p(y_{t-t_k+1:t} | \theta)$ as:
\begin{align}
    \hat{p}_{N_x}(y_{t-t_k+1:t} | \theta)  = \prod_{s=t-t_k+1}^t \left( \frac{1}{N_x} \sum_{i=1}^{N_x} w_s^{i} (x_{s-1}^{i}, x_s^{i}) \right),
\end{align}
and the acceptance probability for the PMMH kernel is given by:
\begin{align}
   \alpha= \min \left\{ 1, \dfrac{\hat{p}_{N_x}(y_{t-t_k+1:t} | \theta^*) p(\theta^*)}{\hat{p}_{N_x}(y_{t-t_k+1:t} | \theta) p(\theta)} \times \dfrac{q(\theta | \theta^*)}{q(\theta^* | \theta)} \right\}.
\end{align}

\section{Posterior estimate in the simulation study}\label{AppC}
\setcounter{table}{0}  
\setcounter{figure}{0}  

Here,  we present the posterior estimates of model parameters from two experimental setups using the O-SMC$^2$ algorithm. The experiments test varying window sizes $t_k$ to evaluate the computational efficiency and the impact of window length on parameter inference.  Table \ref{TableC1} summarizes the parameter settings and prior distributions used in the two simulation experiments.  For each window size, we report the posterior median and 95\% credible intervals, comparing them to SMC$^2$ estimates to assess both inference accuracy and computational savings. Tables \ref{TableC2} and \ref{TableC3} display these results for Experiments 1 and 2, respectively, including the ``Relative Efficiency'' defined as the ratio of the SMC$^2$ computational time to the O-SMC$^2$ computational time. This metric quantifies the time reduction achieved by O-SMC$^2$  for each $t_k$ compared to SMC$^2$, with larger ratios indicating greater computational efficiency.

\begin{table}[H]
\centering
\small
\caption{\footnotesize Summary of parameter settings used in the two simulation Experiments. $\mathcal{U}$(inf, sup) indicates a uniform distribution and  $\mathcal{TN}$ stands for truncated normal distribution ($\mathcal{TN}_{[\text{inf, sup}]}(\text{mean, std})$).}
\resizebox{\textwidth}{!}{%
\begin{tabular}{c|l|l|l|l}
\hline
\textbf{Parameter} & \textbf{Description} & \textbf{Value in simulation} & \textbf{Prior} & \textbf{Notes} \\
\hline\hline
\multicolumn{5}{c}{\textbf{Experiment 1}} \\
\hline
$\beta$ & Transmission rate & $0.6$ & $\mathcal{U}(0, 1)$ & Estimated \\
$\sigma$ & Rate from exposed to infectious & $1/3$ & $\mathcal{U}(0, 1)$ & Estimated \\
$\gamma$ & Recovery rate & $1/5$ & $\mathcal{U}(0, 1)$ & Estimated \\
$I_0$ & Initial infected individuals & 1 & $\mathcal{U}(\{0,\dots,5\})$ & -- \\
$N$ & Population size & $6000$ & -- & Fixed \\
\hline
\multicolumn{5}{c}{\textbf{Experiment 2}} \\
\hline
$\beta(t)$ & Time-varying transmission rate & $\beta(t) = 0.3 \exp\left(\sin\left(\frac{2 \pi t}{55}\right) - \frac{t}{80}\right)$ & Modeled via Eq.~\eqref{rw} & Estimated \\
$\sigma$ & Rate from exposed to infectious & $1/2$ & $\mathcal{U}(0.3, 0.6)$ & Estimated \\
$\gamma$ & Recovery rate & $1/7$ & $\mathcal{TN}_{[0, 1]}(0.14, 0.01)$ & Estimated \\
$\nu_{\beta}$ & Volatility in $\beta(t)$ & -- & $\mathcal{TN}_{[0.05, 0.15]}(0.1, 0.05)$ & Estimated \\
$\beta(0)$ & Initial transmission rate & 0.3& $\mathcal{U}(0.2, 0.5)$ & -- \\
$I_0$ & Initial infected individuals & $5$ & $\mathcal{U}(\{0,\dots,5\})$ &-- \\
$N$ & Population size & $200,000$ & -- & Fixed \\
\hline
\end{tabular}%
}
\label{TableC1}
\end{table}

\begin{table}[H]
\small
\centering
\caption{\footnotesize Summary of parameter estimates in Experiment 1. Posterior medians and 95\% credible intervals were obtained using O-SMC$^2$ with varying window sizes $t_k$ at the final time step $T = 100$. ``Relative Efficiency'' denotes the factor by which computational time is reduced compared to SMC$^2$.}
\resizebox{\textwidth}{!}{%
\begin{tabular}{c|c|c|c|c}
\hline
$t_k$ & $\beta$ & $\sigma$ & $\gamma$ & \text{Relative Efficiency} \\
\hline\hline
20 & 0.611 (0.4751, 0.828) & 0.389 (0.225, 0.652) & 0.209 (0.135, 0.308) & 1.70 \\
40 & 0.603 (0.442, 0.870) & 0.367 (0.219, 0.562) & 0.213 (0.127, 0.338) & 1.18 \\
60 & 0.639 (0.499, 0.847) & 0.334 (0.235, 0.447) & 0.228 (0.158, 0.329) & 1.10 \\
80 & 0.638 (0.496, 0.845) & 0.336 (0.235, 0.479) & 0.224 (0.158, 0.325) & 1.02 \\
SMC$^2$ & 0.612 (0.470, 0.809) & 0.325 (0.236, 0.468) & 0.208 (0.148, 0.305) & - \\
\hline
Truth & 0.6 & 0.333 & 0.2 & - \\
\hline
\end{tabular}
}
\label{TableC2}
\end{table}

\begin{table}[H]
\small
\centering
\caption{\footnotesize Summary of parameter estimates in Experiment 2. Posterior medians and 95\% credible intervals were obtained using O-SMC$^2$ with varying window sizes $t_k$ at the final time step $T = 100$. ``Relative Efficiency'' denotes the factor by which computational time is reduced compared to SMC$^2$.}
\resizebox{\textwidth}{!}{%
\begin{tabular}{c|c|c|c|c}
\hline
$t_k$ & $\sigma$ & $\gamma$ & $\nu_{\beta}$ & Relative Efficiency\\
\hline\hline
20 & 0.477 (0.391, 0.574) & 0.141 (0.130, 0.156) & 0.136 (0.122, 0.148) & 3.22 \\
40 & 0.479 (0.391, 0.573) & 0.141 (0.128, 0.154) & 0.138 (0.125, 0.149) & 1.59 \\
60 & 0.476 (0.393, 0.571) & 0.141 (0.132, 0.150) & 0.139 (0.126, 0.149) & 1.29 \\
80 & 0.475 (0.387, 0.576) & 0.142 (0.131, 0.154) & 0.140 (0.127, 0.149) & 1.06 \\
SMC$^2$ & 0.491 (0.406, 0.588) & 0.143 (0.134, 0.153) & 0.139 (0.127, 0.149) & - \\
\hline
Truth & 0.5 & 0.142 & - &- \\
\hline
\end{tabular}
}
\label{TableC3}
\end{table}

\begin{figure}[H]
	        \centering
       \includegraphics[scale=1]{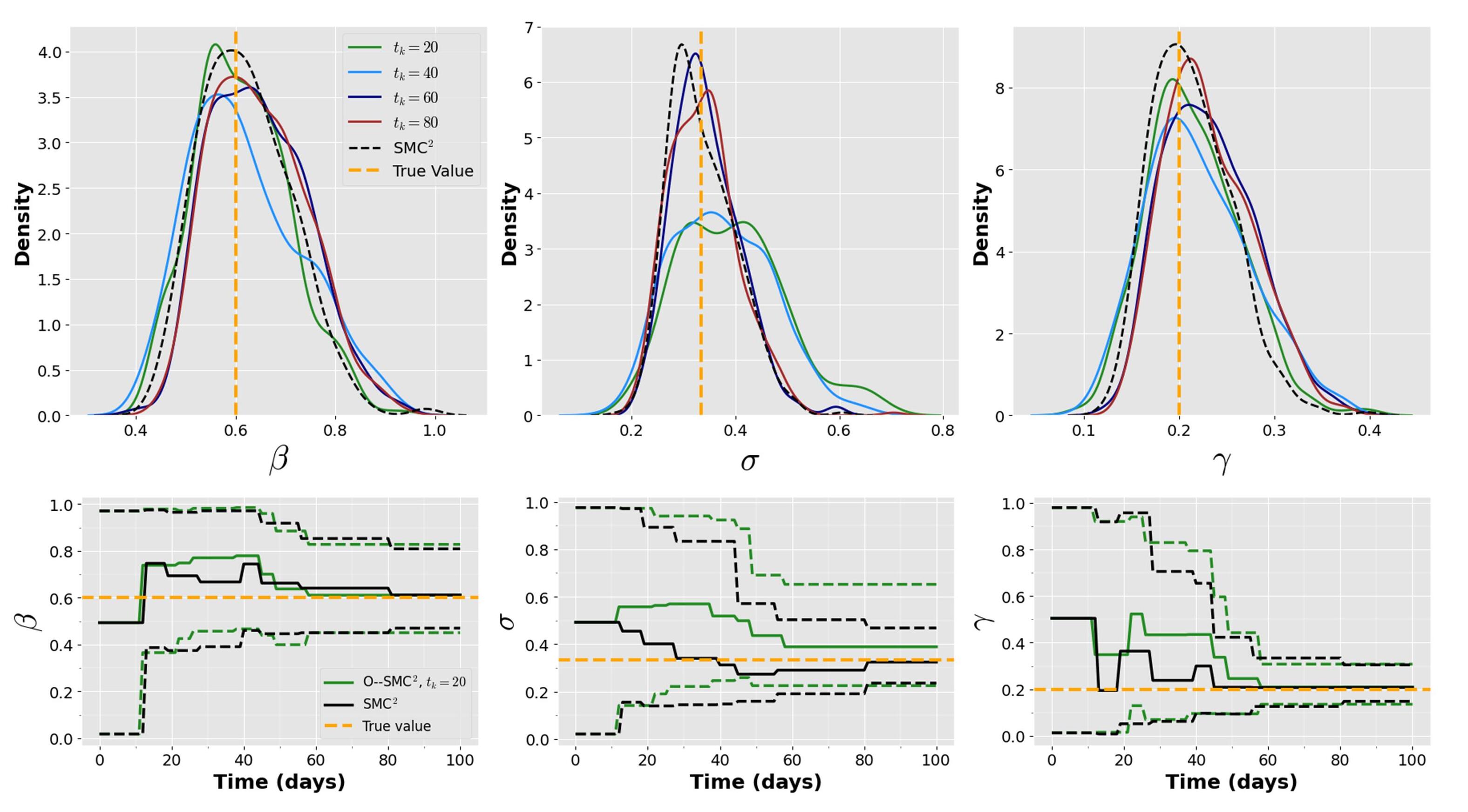}
\caption{\footnotesize Posterior densities of parameters at $T = 100$ (top) and filtered estimates (bottom) in Experiment 1. The dashed lines in the bottom plot indicate the 95\% credible intervals.  True parameter values are represented by orange dashed lines.}
\label{figure:FigC1}
\end{figure}

\begin{figure}[H]
        \centering

 \includegraphics[scale=1]{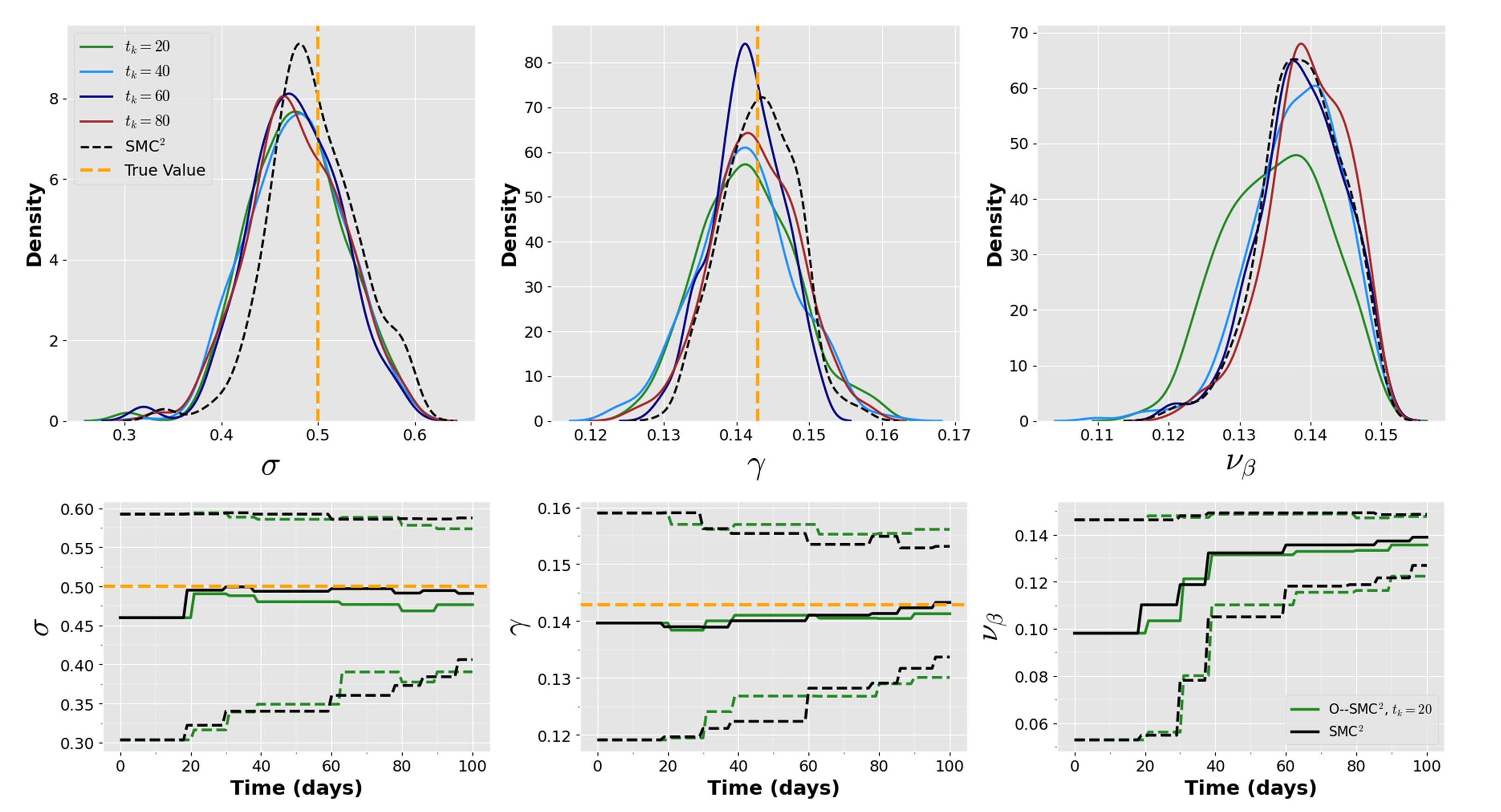}
  
\caption{\footnotesize Posterior densities of parameters at $T = 100$ (top) and filtered estimates (bottom) in Experiment 2.  The dashed lines in the bottom plot indicate the 95\% credible intervals.  True parameter values are represented by orange dashed lines.}
\label{figure:FigC2}

\end{figure}

While smaller $t_k$ values yield significant computational savings, this efficiency advantage decreases as $t_k$ increases. For most tested window sizes, the credible intervals align closely with those from SMC$^2$, indicating minimal bias in this application. However, smaller windows may introduce minor biases if the window does not capture sufficient temporal dependencies, potentially resulting in slightly larger 95\% credible intervals (Figures \ref{figure:FigC1}-\ref{figure:FigC2}).

\begin{figure}[H]
            \centering
       \includegraphics[scale=1]{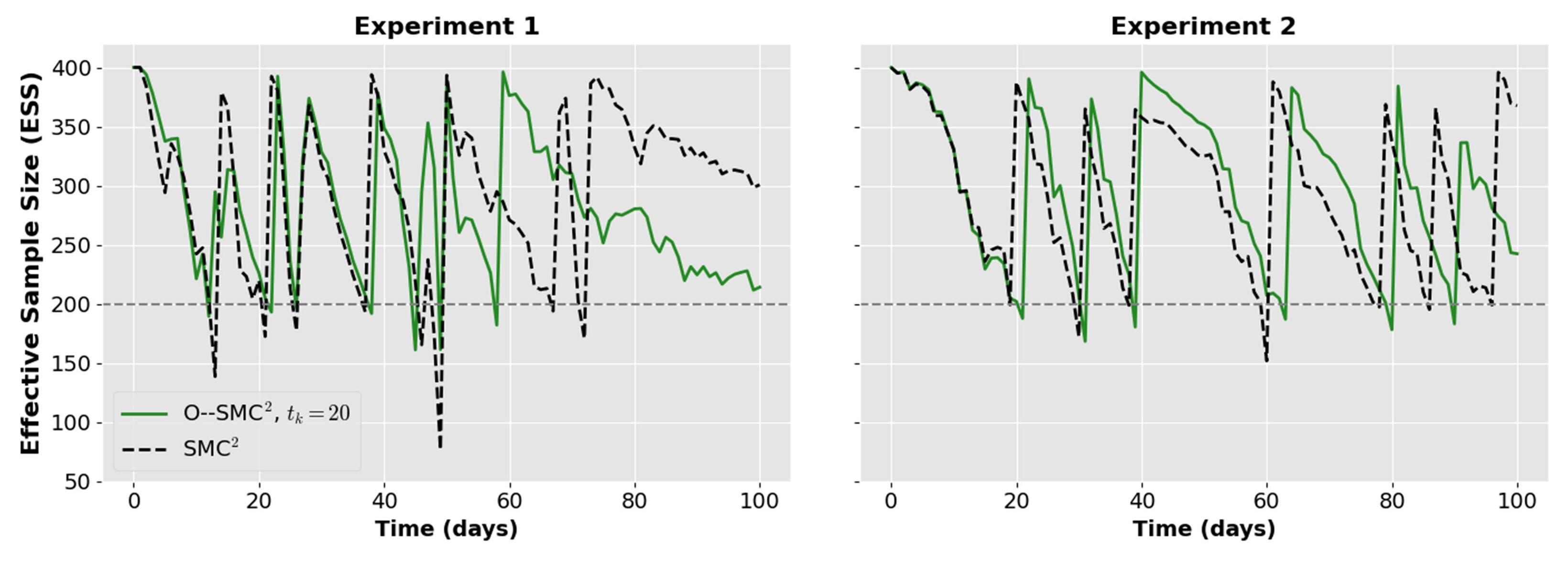}
\caption{\footnotesize  Effective Sample Size. The left graph shows the the evolution of the ESS in Experiment 1 and the right one the evolution in Experiment 2.}
    \label{figure:FigC3}
\end{figure}

It is important to note that the windowing mechanism is applied only during the resampling step (i.e., when ESS drops below 50\% of the total number of parameter particles in our case). Thus, the greater the frequency of resampling at time $t$ (especially when $t$ is significantly larger than $t_k$), the more computationally beneficial O-SMC$^2$ becomes. In Figure \ref{figure:FigC3}, we observe that the frequency of resampling at times well beyond $t_k = 20$ gradually decreases in Experiment 1, while it remains more frequent in Experiment 2. This explains the greater computational gain observed in Experiment 2.

\section{ Parameter settings in the COVID-19 model }\label{AppD}
\setcounter{table}{0}  
\setcounter{figure}{0}  

Table \ref{TableD1} summarizes parameters from the COVID-19 model (Equation \eqref{covm}), including hyperparameters (Equations \eqref{rwm} and \eqref{ov}). For each parameter, prior and posterior estimates (median and 95\% credible interval) are provided. Priors were informed by COVID-19 dynamics literature in Ireland \citep{IEMAG2020, cazelles2021dynamics}.

\begin{table}[H]
\footnotesize
\centering
\caption{\footnotesize Description of the different parameters, priors and posteriors estimate at the final time $T=365$ days. Upper bound and/or lower bound have been imposed by the observations. $\mathcal{TN}$ stands for truncated normal distribution ($\mathcal{TN}_{[\text{inf, sup}]}(\text{mean, std})$).}
\resizebox{\textwidth}{!}{%
\begin{tabular}{c|l|p{3.5cm}|p{3cm}}
\hline
\textbf{Parameter} & \textbf{Description} & \textbf{Prior value} & \textbf{Posterior median (95\% CrI)} \\
\hline\hline
$r_{A}$ & Reduction factor of transmission from $A_{t}$ &  $\mathcal{U}(0.1,0.5)$ & 0.370 (0.304, 0.445) \\
$p_{A}$ & Fraction of infected asymptomatic cases & $\mathcal{U}(0.3,1)$ & 0.413 (0.313, 0.491) \\
$\beta(0)$ & Initial condition transmission rate & $\mathcal{U}(0.6, 0.8)$ & - \\
$\sigma$ & Rate from exposed to infectious & $\mathcal{TN}_{[1/3, 1/5]}(1/4, 0.1)$ & 0.286 (0.256, 0.307) \\
$\gamma$ & Removal rate & $\mathcal{TN}_{[1/7.5, 1/4.5]}(1/6, 0.2)$ & 0.184 (0.165, 0.213) \\
$\nu_{\beta}$ & Volatility of the Brownian process & $\mathcal{TN}_{[0.05, 15]}(0.1, 0.05)$ & 0.137 (0.118, 0.149) \\
$\phi$ & Overdispersion & $\mathcal{U}(0.01, 0.2)$ & 0.100 (0.088, 0.114) \\
\hline
\end{tabular}
}
\label{TableD1}
\end{table}

To evaluate predictive accuracy, we report the Mean Absolute Error (MAE) and the Continuous Ranked Probability Score (CRPS).

The MAE quantifies the average absolute deviation between the observed values and the model’s predictive mean:
\begin{align}
   \text{MAE} = \frac{1}{T} \sum_{t=1}^{T} |y_t - \widehat{\Delta}_t|
\end{align}
where $ y_t $ is the true value of the incidence at time $ t $, and $ \widehat{\Delta}_{t} = \mathrm{E}[\Delta_{t}|y_{1:t}] $ is the model posterior predictive mean of the incidence.

The CRPS evaluates the quality of the full predictive distribution by comparing the predicted cumulative distribution function (CDF) to the empirical CDF of the observation:
\begin{align*}
    \text{CRPS} = \frac{1}{T} \sum_{t=1}^{T} \int_{-\infty}^{\infty} \left( F_t(u) - \mathbb{I}(y_t \leq u) \right)^2 \, du
\end{align*}
where $F_t(u) = p(\Delta_t \leq u \mid y_{1:T})$ is the predictive CDF at time $t$, and $\mathbb{I}(\cdot)$ is the indicator function.

When the predictive distribution is represented by a set of particles $\{x_{t}^i\}_{i=1}^{N_x}$, the CRPS can be approximated as:
\begin{align*}
  \text{CRPS} \approx \frac{1}{T} \sum_{t=1}^{T} \left( \frac{1}{N_x} \sum_{i=1}^{N_x} |y_t - x_{t}^i| - \frac{1}{2N_x^2} \sum_{i=1}^{N_x} \sum_{j=1}^{N_x} |x_{t}^i - x_{t}^j| \right)
\end{align*}

\begin{figure}[H]
	    \centering
      \includegraphics[scale=1]{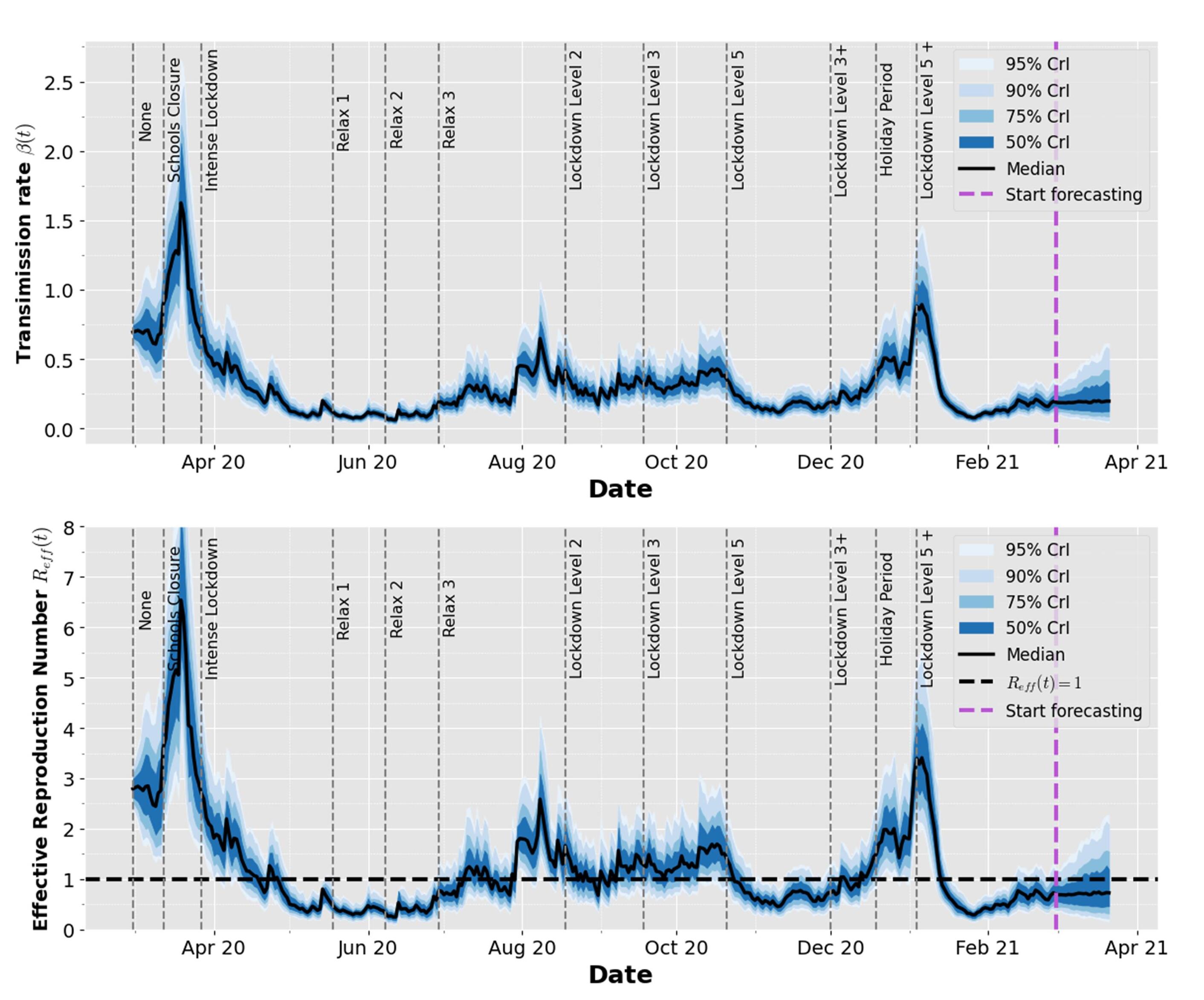}
	\caption{\footnotesize  Daily estimates of the time-dependent transmission rate 
 and effective reproduction number. The solid black lines represent the median estimate of $R_{\text{eff}}(t)$, with shaded blue regions indicating posterior uncertainty quantiles. Weekly averaged estimates of $R_{\text{eff}}(t)$ are shown in Figure \ref{figure:Fig6}.}
 \label{figure:FigD1}
\end{figure}

\section{Application in real-time surveillance}\label{AppE}
\setcounter{table}{0}  
\setcounter{figure}{0}  

\begin{figure}[H]
        \centering
 \includegraphics[scale=1]{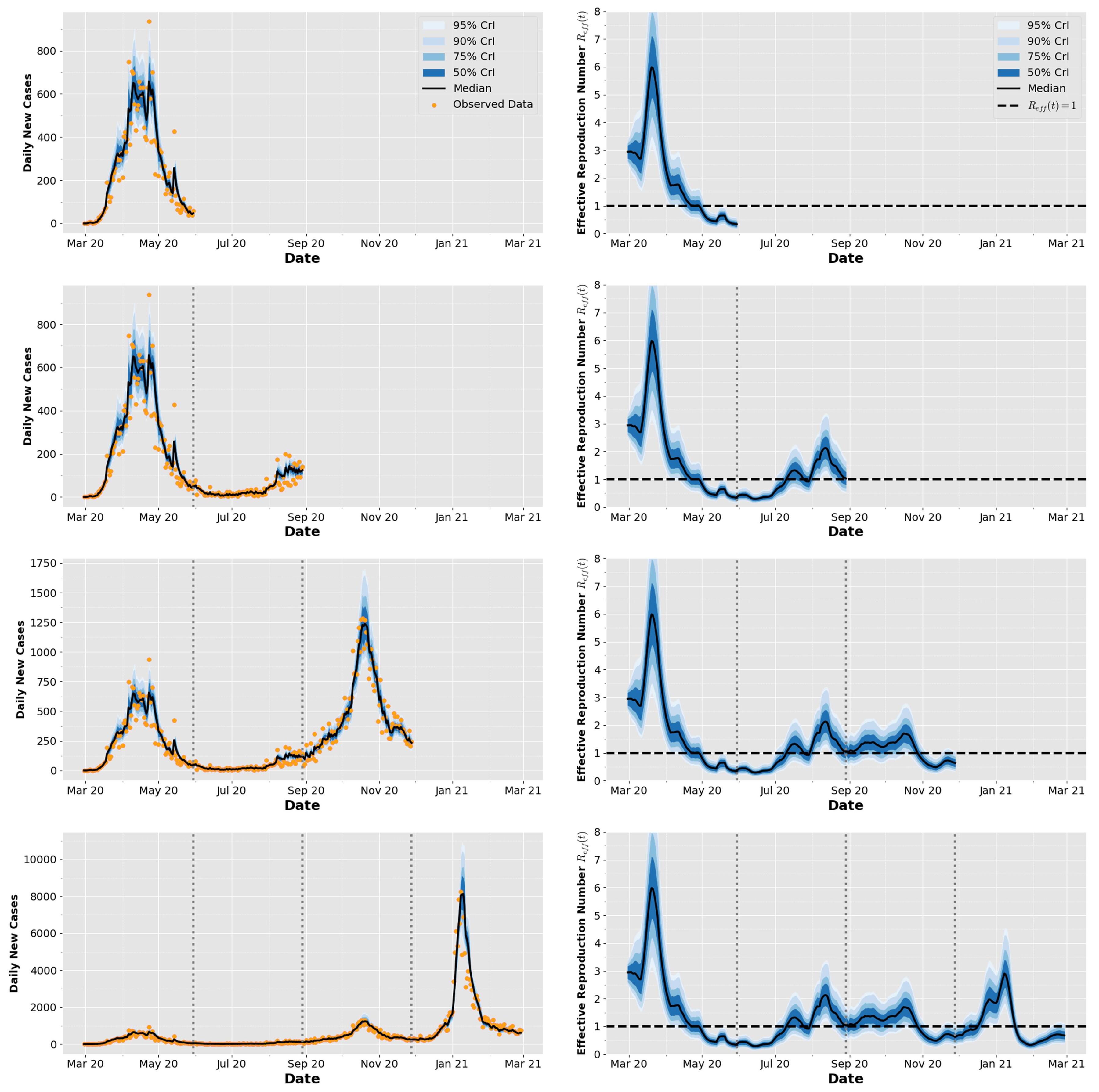}
\caption{\footnotesize  Real-time application of O-SMC$^2$ to sequential data: (Row 1) February 29, 2020, to May 30, 2020; (Row 2) May 30, 2020, to August 29, 2020; (Row 3) August 29, 2020, to November 28, 2020; and (Row 4) November 28, 2020, to February 27, 2021. Each row depicts the sequential estimates of daily new cases and the weekly average effective reproduction number ($R_{\text{eff}}(t)$). The solid black lines represent the median estimate, while the shaded blue regions indicate posterior uncertainty quantiles.}
    \label{figure:FigE1}
\end{figure}

The methodology developed in this paper is applied to a real-time surveillance setting, demonstrating how our model can track evolving dynamics in sequential data. By integrating new observations over time, the model provides insights into the system’s behavior. The focus is on an adaptive framework that updates the posterior distribution, conditioned on incoming data, while utilizing results from previous O-SMC$^2$ runs as prior information. The results are shown in Figure \ref{figure:FigE1}, where we observe that there is no significant difference compared to the results obtained from running the model from scratch (see Figure \ref{figure:Fig6}).

\begin{thebibliography}{47}
\expandafter\ifx\csname natexlab\endcsname\relax\def\natexlab#1{#1}\fi
\providecommand{\url}[1]{\texttt{#1}}
\providecommand{\href}[2]{#2}
\providecommand{\path}[1]{#1}
\providecommand{\DOIprefix}{doi:}
\providecommand{\ArXivprefix}{arXiv:}
\providecommand{\URLprefix}{URL: }
\providecommand{\Pubmedprefix}{pmid:}
\providecommand{\doi}[1]{\href{http://dx.doi.org/#1}{\path{#1}}}
\providecommand{\Pubmed}[1]{\href{pmid:#1}{\path{#1}}}
\providecommand{\bibinfo}[2]{#2}
\ifx\xfnm\relax \def\xfnm[#1]{\unskip,\space#1}\fi
\bibitem[{Aloke et~al.(2023)Aloke, Nwaeze, Omenyi and Uchenna}]{aloke2023parameters}
\bibinfo{author}{Aloke, S.N.}, \bibinfo{author}{Nwaeze, E.}, \bibinfo{author}{Omenyi, L.}, \bibinfo{author}{Uchenna, M.}, \bibinfo{year}{2023}.
\newblock \bibinfo{title}{Parameters estimation of covid-19 seir model}.
\newblock \bibinfo{journal}{Asian Journal of Pure and Applied Mathematics} , \bibinfo{pages}{229--241}.
\bibitem[{Althaus(2014)}]{althaus2014estimating}
\bibinfo{author}{Althaus, C.L.}, \bibinfo{year}{2014}.
\newblock \bibinfo{title}{{Estimating the reproduction number of Ebola virus (EBOV) during the 2014 outbreak in West Africa}}.
\newblock \bibinfo{journal}{PLoS currents} \bibinfo{volume}{6}, \bibinfo{pages}{ecurrents--outbreaks}.
\bibitem[{Andrieu et~al.(2010)Andrieu, Doucet and Holenstein}]{Andrieu2010}
\bibinfo{author}{Andrieu, C.}, \bibinfo{author}{Doucet, A.}, \bibinfo{author}{Holenstein, R.}, \bibinfo{year}{2010}.
\newblock \bibinfo{title}{{Particle Markov Chain Monte Carlo Methods (with discussion)}}.
\newblock \bibinfo{journal}{Journal of the Royal Statistical Society: Series B (Statistical Methodology)} \bibinfo{volume}{72}, \bibinfo{pages}{1--269}.
\bibitem[{Birrell et~al.(2017)Birrell, Pebody, Charlett, Zhang and Angelis}]{Birrell2017}
\bibinfo{author}{Birrell, P.J.}, \bibinfo{author}{Pebody, R.G.}, \bibinfo{author}{Charlett, A.}, \bibinfo{author}{Zhang, X.S.}, \bibinfo{author}{Angelis, D.D.}, \bibinfo{year}{2017}.
\newblock \bibinfo{title}{Real-time modelling of a pandemic influenza outbreak}.
\newblock \bibinfo{journal}{Health Technology Assessment} \bibinfo{volume}{21}, \bibinfo{pages}{1--118}.
\newblock \DOIprefix\doi{10.3310/hta21580}.
\bibitem[{Birrell et~al.(2020)Birrell, Wernisch, Tom, Held, Roberts, Pebody and De~Angelis}]{birrell2020efficient}
\bibinfo{author}{Birrell, P.J.}, \bibinfo{author}{Wernisch, L.}, \bibinfo{author}{Tom, B.D.}, \bibinfo{author}{Held, L.}, \bibinfo{author}{Roberts, G.O.}, \bibinfo{author}{Pebody, R.G.}, \bibinfo{author}{De~Angelis, D.}, \bibinfo{year}{2020}.
\newblock \bibinfo{title}{{Efficient real-time monitoring of an emerging influenza pandemic: How feasible?}}
\newblock \bibinfo{journal}{The annals of applied statistics} \bibinfo{volume}{14}, \bibinfo{pages}{74}.
\bibitem[{Calvetti et~al.(2021)Calvetti, Hoover, Rose and Somersalo}]{calvetti2021}
\bibinfo{author}{Calvetti, D.}, \bibinfo{author}{Hoover, A.}, \bibinfo{author}{Rose, J.}, \bibinfo{author}{Somersalo, E.}, \bibinfo{year}{2021}.
\newblock \bibinfo{title}{Bayesian particle filter algorithm for learning epidemic dynamics}.
\newblock \bibinfo{journal}{Inverse Problems} \bibinfo{volume}{37}, \bibinfo{pages}{Article 115008}.
\newblock \DOIprefix\doi{10.1088/1361-6420/ac2cdc}.
\bibitem[{Camacho et~al.(2015)Camacho, Kucharski, A~Ki-Sawyerr, White, Flasche, Bagguelin, Pollington, Carney, Glover et~al.}]{camacho2015temporal}
\bibinfo{author}{Camacho, A.}, \bibinfo{author}{Kucharski, A.}, \bibinfo{author}{A~Ki-Sawyerr, Y.}, \bibinfo{author}{White, M.A.}, \bibinfo{author}{Flasche, S.}, \bibinfo{author}{Bagguelin, M.}, \bibinfo{author}{Pollington, T.}, \bibinfo{author}{Carney, J.R.}, \bibinfo{author}{Glover, R.}, et~al., \bibinfo{year}{2015}.
\newblock \bibinfo{title}{{Temporal changes in Ebola transmission in Sierra Leone and implications for control requirements: A real-time modelling study}}.
\newblock \bibinfo{journal}{PLoS Curr} \bibinfo{volume}{7}.
\bibitem[{Carpenter et~al.(2017)Carpenter, Gelman, Hoffman, Lee, Goodrich, Betancourt, Brubaker, Guo, Li and Riddell}]{carpenter2017stan}
\bibinfo{author}{Carpenter, B.}, \bibinfo{author}{Gelman, A.}, \bibinfo{author}{Hoffman, M.D.}, \bibinfo{author}{Lee, D.}, \bibinfo{author}{Goodrich, B.}, \bibinfo{author}{Betancourt, M.}, \bibinfo{author}{Brubaker, M.}, \bibinfo{author}{Guo, J.}, \bibinfo{author}{Li, P.}, \bibinfo{author}{Riddell, A.}, \bibinfo{year}{2017}.
\newblock \bibinfo{title}{Stan: A probabilistic programming language}.
\newblock \bibinfo{journal}{Journal of statistical software} \bibinfo{volume}{76}, \bibinfo{pages}{1--32}.
\bibitem[{Cazelles et~al.(2021)Cazelles, Nguyen-Van-Yen, Champagne et~al.}]{cazelles2021dynamics}
\bibinfo{author}{Cazelles, B.}, \bibinfo{author}{Nguyen-Van-Yen, B.}, \bibinfo{author}{Champagne, C.}, et~al., \bibinfo{year}{2021}.
\newblock \bibinfo{title}{{Dynamics of the COVID-19 epidemic in Ireland under mitigation}}.
\newblock \bibinfo{journal}{BMC Infectious Diseases} \bibinfo{volume}{21}, \bibinfo{pages}{735}.
\newblock \URLprefix \url{https://doi.org/10.1186/s12879-021-06433-9}, \DOIprefix\doi{10.1186/s12879-021-06433-9}.
\bibitem[{Chopin(2002)}]{chopin2002sequential}
\bibinfo{author}{Chopin, N.}, \bibinfo{year}{2002}.
\newblock \bibinfo{title}{A sequential particle filter method for static models}.
\newblock \bibinfo{journal}{Biometrika} \bibinfo{volume}{89}, \bibinfo{pages}{539--552}.
\bibitem[{Chopin et~al.(2013)Chopin, Jacob and Papaspiliopoulos}]{Chopin2013}
\bibinfo{author}{Chopin, N.}, \bibinfo{author}{Jacob, P.E.}, \bibinfo{author}{Papaspiliopoulos, O.}, \bibinfo{year}{2013}.
\newblock \bibinfo{title}{{SMC$^2$: An Efficient Algorithm for Sequential Analysis of State Space Models}}.
\newblock \bibinfo{journal}{Journal of the Royal Statistical Society: Series B (Statistical Methodology)} \bibinfo{volume}{75}, \bibinfo{pages}{397--426}.
\bibitem[{Chowell et~al.(2024)Chowell, Bleichrodt and Luo}]{chowell2024parameter}
\bibinfo{author}{Chowell, G.}, \bibinfo{author}{Bleichrodt, A.}, \bibinfo{author}{Luo, R.}, \bibinfo{year}{2024}.
\newblock \bibinfo{title}{Parameter estimation and forecasting with quantified uncertainty for ordinary differential equation models using quantdiffforecast: A matlab toolbox and tutorial}.
\newblock \bibinfo{journal}{Statistics in Medicine} \bibinfo{volume}{43}, \bibinfo{pages}{1826--1848}.
\bibitem[{Dong et~al.(2023)Dong, Goodbrake, Harrington and Pogudin}]{dong2023differential}
\bibinfo{author}{Dong, R.}, \bibinfo{author}{Goodbrake, C.}, \bibinfo{author}{Harrington, H.A.}, \bibinfo{author}{Pogudin, G.}, \bibinfo{year}{2023}.
\newblock \bibinfo{title}{Differential elimination for dynamical models via projections with applications to structural identifiability}.
\newblock \bibinfo{journal}{SIAM Journal on Applied Algebra and Geometry} \bibinfo{volume}{7}, \bibinfo{pages}{194--235}.
\bibitem[{Doucet et~al.(2001)Doucet, de~Freitas and Gordon}]{Doucet2001}
\bibinfo{author}{Doucet, A.}, \bibinfo{author}{de~Freitas, N.}, \bibinfo{author}{Gordon, N.}, \bibinfo{year}{2001}.
\newblock \bibinfo{title}{{An Introduction to Sequential Monte Carlo Methods}}.
\newblock \bibinfo{publisher}{Springer New York}, \bibinfo{address}{New York, NY}.
\newblock \DOIprefix\doi{10.1007/978-1-4757-3437-9_1}.
\bibitem[{Doucet et~al.(2009)Doucet, Johansen et~al.}]{doucet2009tutorial}
\bibinfo{author}{Doucet, A.}, \bibinfo{author}{Johansen, A.M.}, et~al., \bibinfo{year}{2009}.
\newblock \bibinfo{title}{{A tutorial on particle filtering and smoothing: Fifteen years later}}.
\newblock \bibinfo{journal}{Handbook of nonlinear filtering} \bibinfo{volume}{12}, \bibinfo{pages}{3}.
\bibitem[{Dureau et~al.(2013)Dureau, Kalogeropoulos and Baguelin}]{Dureau2013}
\bibinfo{author}{Dureau, J.}, \bibinfo{author}{Kalogeropoulos, K.}, \bibinfo{author}{Baguelin, M.}, \bibinfo{year}{2013}.
\newblock \bibinfo{title}{Capturing the time-varying drivers of an epidemic using stochastic dynamical systems}.
\newblock \bibinfo{journal}{Biostatistics} \bibinfo{volume}{14}, \bibinfo{pages}{541--555}.
\newblock \DOIprefix\doi{10.1093/biostatistics/kxs052}, \Pubmedprefix \Pubmed{23292757}. \bibinfo{note}{epub 2013 Jan 4}.
\bibitem[{Endo et~al.(2019)Endo, van Leeuwen and Baguelin}]{endo2019introduction}
\bibinfo{author}{Endo, A.}, \bibinfo{author}{van Leeuwen, E.}, \bibinfo{author}{Baguelin, M.}, \bibinfo{year}{2019}.
\newblock \bibinfo{title}{{Introduction to particle Markov-chain Monte Carlo for disease dynamics modellers}}.
\newblock \bibinfo{journal}{Epidemics} \bibinfo{volume}{29}, \bibinfo{pages}{100363}.
\bibitem[{Funk et~al.(2018)Funk, Camacho, Kucharski, Eggo and Edmunds}]{funk2018realtime}
\bibinfo{author}{Funk, S.}, \bibinfo{author}{Camacho, A.}, \bibinfo{author}{Kucharski, A.J.}, \bibinfo{author}{Eggo, R.M.}, \bibinfo{author}{Edmunds, W.J.}, \bibinfo{year}{2018}.
\newblock \bibinfo{title}{{Real-time forecasting of infectious disease dynamics with a stochastic semi-mechanistic model}}.
\newblock \bibinfo{journal}{Epidemics} \bibinfo{volume}{22}, \bibinfo{pages}{56--61}.
\newblock \DOIprefix\doi{10.1016/j.epidem.2016.11.003}.
\bibitem[{Golightly and Kypraios(2018)}]{Golightly2018}
\bibinfo{author}{Golightly, A.}, \bibinfo{author}{Kypraios, T.}, \bibinfo{year}{2018}.
\newblock \bibinfo{title}{Efficient smc 2 schemes for stochastic kinetic models}.
\newblock \bibinfo{journal}{Statistics and Computing} \bibinfo{volume}{28}, \bibinfo{pages}{1215--1230}.
\bibitem[{Gordon et~al.(1993)Gordon, Salmond and Smith}]{gordon1993novel}
\bibinfo{author}{Gordon, N.}, \bibinfo{author}{Salmond, D.}, \bibinfo{author}{Smith, A.}, \bibinfo{year}{1993}.
\newblock \bibinfo{title}{{Novel Approach to Nonlinear/Non-Gaussian Bayesian State Estimation}}.
\newblock \bibinfo{journal}{IEEE Proceedings F – Radar and Signal Processing} \bibinfo{volume}{140}, \bibinfo{pages}{107--113}.
\newblock \DOIprefix\doi{10.1049/ip-f-2.1993.0015}.
\bibitem[{Grinsztajn et~al.(2021)Grinsztajn, Semenova, Margossian and Riou}]{grinsztajn2021bayesian}
\bibinfo{author}{Grinsztajn, L.}, \bibinfo{author}{Semenova, E.}, \bibinfo{author}{Margossian, C.C.}, \bibinfo{author}{Riou, J.}, \bibinfo{year}{2021}.
\newblock \bibinfo{title}{Bayesian workflow for disease transmission modeling in stan}.
\newblock \bibinfo{journal}{Statistics in medicine} \bibinfo{volume}{40}, \bibinfo{pages}{6209--6234}.
\bibitem[{Han et~al.(2025)Han, Kim, Koh, Kobayashi and Choi}]{han2025sequential}
\bibinfo{author}{Han, D.}, \bibinfo{author}{Kim, M.}, \bibinfo{author}{Koh, E.}, \bibinfo{author}{Kobayashi, G.}, \bibinfo{author}{Choi, T.}, \bibinfo{year}{2025}.
\newblock \bibinfo{title}{Sequential monte carlo abc: an overview with application to covid-19 data}.
\newblock \bibinfo{journal}{Journal of the Korean Statistical Society} \bibinfo{volume}{54}, \bibinfo{pages}{248--283}.
\bibitem[{Inouzhe et~al.(2023)Inouzhe, Rodríguez-Álvarez, Nagar and Akhmatskaya}]{inouzhe2023}
\bibinfo{author}{Inouzhe, H.}, \bibinfo{author}{Rodríguez-Álvarez, M.X.}, \bibinfo{author}{Nagar, L.}, \bibinfo{author}{Akhmatskaya, E.}, \bibinfo{year}{2023}.
\newblock \bibinfo{title}{{Dynamic SIR/SEIR-like models comprising a time-dependent transmission rate: Hamiltonian Monte Carlo approach with applications to COVID-19}}.
\newblock \URLprefix \url{https://arxiv.org/abs/2301.06385}, \href{http://arxiv.org/abs/2301.06385}{{\tt arXiv:2301.06385}}.
\bibitem[{{Irish Department of Health}(2020)}]{IEMAG2020}
\bibinfo{author}{{Irish Department of Health}}, \bibinfo{year}{2020}.
\newblock \bibinfo{title}{{A Population-Level SEIR Model for COVID-19 Scenarios}}.
\newblock \bibinfo{howpublished}{Available from: \url{www.hse.ie}}.
\bibitem[{Jacob(2015)}]{Jacob2015}
\bibinfo{author}{Jacob, P.E.}, \bibinfo{year}{2015}.
\newblock \bibinfo{title}{{Sequential Bayesian inference for implicit hidden Markov models and current limitations}}.
\newblock \bibinfo{journal}{ESAIM: Proceedings and Surveys} \bibinfo{volume}{51}, \bibinfo{pages}{24--48}.
\bibitem[{Jaouimaa et~al.(2021)Jaouimaa, Dempsey, Van~Osch, Kinsella, Burke, Wyse and Sweeney}]{jaouimaa2021age}
\bibinfo{author}{Jaouimaa, F.Z.}, \bibinfo{author}{Dempsey, D.}, \bibinfo{author}{Van~Osch, S.}, \bibinfo{author}{Kinsella, S.}, \bibinfo{author}{Burke, K.}, \bibinfo{author}{Wyse, J.}, \bibinfo{author}{Sweeney, J.}, \bibinfo{year}{2021}.
\newblock \bibinfo{title}{{An age-structured SEIR model for COVID-19 incidence in Dublin, Ireland with framework for evaluating health intervention cost}}.
\newblock \bibinfo{journal}{PLoS One} \bibinfo{volume}{16}, \bibinfo{pages}{e0260632}.
\newblock \DOIprefix\doi{10.1371/journal.pone.0260632}.
\bibitem[{Jewell et~al.(2009)Jewell, Kypraios, Neal and Roberts}]{jewell2009bayesian}
\bibinfo{author}{Jewell, C.P.}, \bibinfo{author}{Kypraios, T.}, \bibinfo{author}{Neal, P.}, \bibinfo{author}{Roberts, G.O.}, \bibinfo{year}{2009}.
\newblock \bibinfo{title}{{Bayesian analysis for emerging infectious diseases}}.
\newblock \bibinfo{journal}{Bayesian Analysis} \bibinfo{volume}{4}, \bibinfo{pages}{465--496}.
\newblock \DOIprefix\doi{10.1214/09-BA417}.
\bibitem[{Kantas et~al.(2015)Kantas, Doucet, Singh, Maciejowski, Chopin et~al.}]{kantas2015particle}
\bibinfo{author}{Kantas, N.}, \bibinfo{author}{Doucet, A.}, \bibinfo{author}{Singh, S.S.}, \bibinfo{author}{Maciejowski, J.}, \bibinfo{author}{Chopin, N.}, et~al., \bibinfo{year}{2015}.
\newblock \bibinfo{title}{On particle methods for parameter estimation in state-space models}.
\newblock \bibinfo{journal}{Statistical science} \bibinfo{volume}{30}, \bibinfo{pages}{328--351}.
\bibitem[{Karami et~al.(2024)Karami, Bleichrodt, Luo and Chowell}]{karami2024bayesianfitforecast}
\bibinfo{author}{Karami, H.}, \bibinfo{author}{Bleichrodt, A.}, \bibinfo{author}{Luo, R.}, \bibinfo{author}{Chowell, G.}, \bibinfo{year}{2024}.
\newblock \bibinfo{title}{Bayesianfitforecast: A user-friendly r toolbox for parameter estimation and forecasting with ordinary differential equations}.
\newblock \bibinfo{journal}{arXiv preprint arXiv:2411.05371} .
\bibitem[{Kypraios et~al.(2017)Kypraios, Neal and Prangle}]{kypraios2017tutorial}
\bibinfo{author}{Kypraios, T.}, \bibinfo{author}{Neal, P.}, \bibinfo{author}{Prangle, D.}, \bibinfo{year}{2017}.
\newblock \bibinfo{title}{{A tutorial introduction to Bayesian inference for stochastic epidemic models using Approximate Bayesian Computation}}.
\newblock \bibinfo{journal}{Mathematical Biosciences} \bibinfo{volume}{287}, \bibinfo{pages}{42--53}.
\bibitem[{Lekone and Finkenst{\"a}dt(2006)}]{lekone2006statistical}
\bibinfo{author}{Lekone, P.E.}, \bibinfo{author}{Finkenst{\"a}dt, B.F.}, \bibinfo{year}{2006}.
\newblock \bibinfo{title}{{Statistical inference in a stochastic epidemic SEIR model with control intervention: Ebola as a case study}}.
\newblock \bibinfo{journal}{Biometrics} \bibinfo{volume}{62}, \bibinfo{pages}{1170--1177}.
\bibitem[{Liyanage et~al.(2025)Liyanage, Saucedo, Tuncer and Chowell}]{liyanage2025tutorial}
\bibinfo{author}{Liyanage, Y.R.}, \bibinfo{author}{Saucedo, O.}, \bibinfo{author}{Tuncer, N.}, \bibinfo{author}{Chowell, G.}, \bibinfo{year}{2025}.
\newblock \bibinfo{title}{A tutorial on structural identifiability of epidemic models using structuralidentifiability. jl}.
\newblock \bibinfo{journal}{arXiv preprint arXiv:2505.10517} .
\bibitem[{Marani et~al.(2021)Marani, Katul, Pan and Parolari}]{marani2021intensity}
\bibinfo{author}{Marani, M.}, \bibinfo{author}{Katul, G.G.}, \bibinfo{author}{Pan, W.K.}, \bibinfo{author}{Parolari, A.J.}, \bibinfo{year}{2021}.
\newblock \bibinfo{title}{Intensity and frequency of extreme novel epidemics}.
\newblock \bibinfo{journal}{Proceedings National Academy Sciences} \bibinfo{volume}{118}.
\newblock \DOIprefix\doi{10.1073/pnas.2105482118}.
\bibitem[{Minter and Retkute(2019)}]{minter2019approximate}
\bibinfo{author}{Minter, A.}, \bibinfo{author}{Retkute, R.}, \bibinfo{year}{2019}.
\newblock \bibinfo{title}{Approximate bayesian computation for infectious disease modelling}.
\newblock \bibinfo{journal}{Epidemics} \bibinfo{volume}{29}, \bibinfo{pages}{100368}.
\bibitem[{O’Neill and Roberts(1999)}]{o1999bayesian}
\bibinfo{author}{O’Neill, P.D.}, \bibinfo{author}{Roberts, G.O.}, \bibinfo{year}{1999}.
\newblock \bibinfo{title}{Bayesian inference for partially observed stochastic epidemics}.
\newblock \bibinfo{journal}{Journal of the Royal Statistical Society Series A: Statistics in Society} \bibinfo{volume}{162}, \bibinfo{pages}{121--129}.
\bibitem[{Pitt and Shephard(1999)}]{pitt1999filtering}
\bibinfo{author}{Pitt, M.K.}, \bibinfo{author}{Shephard, N.}, \bibinfo{year}{1999}.
\newblock \bibinfo{title}{{Filtering via Simulation: Auxiliary Particle Filters}}.
\newblock \bibinfo{journal}{Journal of the American Statistical Association} \bibinfo{volume}{94}, \bibinfo{pages}{590--599}.
\newblock \DOIprefix\doi{10.1080/01621459.1999.10474153}.
\bibitem[{Roberts et~al.(2015)Roberts, Andreasen, Lloyd and Pellis}]{roberts2015nine}
\bibinfo{author}{Roberts, M.}, \bibinfo{author}{Andreasen, V.}, \bibinfo{author}{Lloyd, A.}, \bibinfo{author}{Pellis, L.}, \bibinfo{year}{2015}.
\newblock \bibinfo{title}{Nine challenges for deterministic epidemic models}.
\newblock \bibinfo{journal}{Epidemics} \bibinfo{volume}{10}, \bibinfo{pages}{49--53}.
\bibitem[{Rosato et~al.(2023)Rosato, Varsi, Murphy and Maskell}]{rosato2023log}
\bibinfo{author}{Rosato, C.}, \bibinfo{author}{Varsi, A.}, \bibinfo{author}{Murphy, J.}, \bibinfo{author}{Maskell, S.}, \bibinfo{year}{2023}.
\newblock \bibinfo{title}{An o (log 2 n) smc 2 algorithm on distributed memory with an approx. optimal l-kernel}, in: \bibinfo{booktitle}{2023 IEEE Symposium Sensor Data Fusion and International Conference on Multisensor Fusion and Integration (SDF-MFI)}, \bibinfo{organization}{IEEE}. pp. \bibinfo{pages}{1--8}.
\bibitem[{Sauer et~al.(2021)Sauer, Berry, Ebeigbe, Norton, Whalen and Schiff}]{sauer2021identifiability}
\bibinfo{author}{Sauer, T.}, \bibinfo{author}{Berry, T.}, \bibinfo{author}{Ebeigbe, D.}, \bibinfo{author}{Norton, M.M.}, \bibinfo{author}{Whalen, A.J.}, \bibinfo{author}{Schiff, S.J.}, \bibinfo{year}{2021}.
\newblock \bibinfo{title}{{Identifiability of infection model parameters early in an epidemic}}.
\newblock \bibinfo{journal}{SIAM journal on control and optimization} \bibinfo{volume}{60}, \bibinfo{pages}{S27--S48}.
\bibitem[{Sheinson et~al.(2014)Sheinson, Niemi and Meiring}]{SHEINSON201421}
\bibinfo{author}{Sheinson, D.M.}, \bibinfo{author}{Niemi, J.}, \bibinfo{author}{Meiring, W.}, \bibinfo{year}{2014}.
\newblock \bibinfo{title}{Comparison of the performance of particle filter algorithms applied to tracking of a disease epidemic}.
\newblock \bibinfo{journal}{Mathematical Biosciences} \bibinfo{volume}{255}, \bibinfo{pages}{21--32}.
\newblock \DOIprefix\doi{https://doi.org/10.1016/j.mbs.2014.06.018}.
\bibitem[{Storvik et~al.(2023)Storvik, Diz-Lois~Palomares, Engebretsen, R{\o}, Eng{\o}-Monsen, Kristoffersen, De~Blasio and Frigessi}]{storvik2023sequential}
\bibinfo{author}{Storvik, G.}, \bibinfo{author}{Diz-Lois~Palomares, A.}, \bibinfo{author}{Engebretsen, S.}, \bibinfo{author}{R{\o}, G.{\O}.I.}, \bibinfo{author}{Eng{\o}-Monsen, K.}, \bibinfo{author}{Kristoffersen, A.B.}, \bibinfo{author}{De~Blasio, B.F.}, \bibinfo{author}{Frigessi, A.}, \bibinfo{year}{2023}.
\newblock \bibinfo{title}{{A sequential Monte Carlo approach to estimate a time-varying reproduction number in infectious disease models: the Covid-19 case}}.
\newblock \bibinfo{journal}{Journal of the Royal Statistical Society Series A: Statistics in Society} \bibinfo{volume}{186}, \bibinfo{pages}{616--632}.
\bibitem[{Toni et~al.(2009)Toni, Welch, Strelkowa, Ipsen and Stumpf}]{toni2009approximate}
\bibinfo{author}{Toni, T.}, \bibinfo{author}{Welch, D.}, \bibinfo{author}{Strelkowa, N.}, \bibinfo{author}{Ipsen, A.}, \bibinfo{author}{Stumpf, M.P.}, \bibinfo{year}{2009}.
\newblock \bibinfo{title}{Approximate bayesian computation scheme for parameter inference and model selection in dynamical systems}.
\newblock \bibinfo{journal}{Journal of the Royal Society Interface} \bibinfo{volume}{6}, \bibinfo{pages}{187--202}.
\bibitem[{Vieira(2018)}]{vieira2018bayesian}
\bibinfo{author}{Vieira, R.M.}, \bibinfo{year}{2018}.
\newblock \bibinfo{title}{Bayesian online state and parameter estimation for streaming data}.
\newblock Ph.D. thesis. Newcastle University.
\bibitem[{Wang and Walker(2022)}]{wang2022bayes}
\bibinfo{author}{Wang, S.}, \bibinfo{author}{Walker, S.G.}, \bibinfo{year}{2022}.
\newblock \bibinfo{title}{{Bayesian Data Augmentation for Partially Observed Stochastic Compartmental Models}}.
\newblock \URLprefix \url{https://arxiv.org/abs/2206.09018}, \href{http://arxiv.org/abs/2206.09018}{{\tt arXiv:2206.09018}}.
\bibitem[{Welding and Neal(2019)}]{welding2019real}
\bibinfo{author}{Welding, J.}, \bibinfo{author}{Neal, P.}, \bibinfo{year}{2019}.
\newblock \bibinfo{title}{Real-time analysis of epidemic data}.
\newblock \bibinfo{journal}{arXiv preprint arXiv:1909.11560} \DOIprefix\doi{10.48550/arXiv.1909.11560}.
\bibitem[{{World Health Organisation (WHO)}(2024)}]{who2020covid}
\bibinfo{author}{{World Health Organisation (WHO)}}, \bibinfo{year}{2024}.
\newblock \bibinfo{title}{{Coronavirus Disease (COVID-19) Pandemic}}.
\newblock \bibinfo{howpublished}{\url{https://www.who.int/emergencies/diseases/novel-coronavirus-2019}}.
\newblock \bibinfo{note}{Last accessed on 2024-03-1}.
\bibitem[{Yang et~al.(2014)Yang, Karspeck and Shaman}]{yang2014comparison}
\bibinfo{author}{Yang, W.}, \bibinfo{author}{Karspeck, A.}, \bibinfo{author}{Shaman, J.}, \bibinfo{year}{2014}.
\newblock \bibinfo{title}{{Comparison of Filtering Methods for the Modeling and Retrospective Forecasting of Influenza Epidemics}}.
\newblock \bibinfo{journal}{PLoS Comput Biol} \bibinfo{volume}{10}, \bibinfo{pages}{e1003583}.
\newblock \DOIprefix\doi{10.1371/journal.pcbi.1003583}.

\end{thebibliography}

\end{document}